\documentclass[11pt,onecolumn]{IEEEtran2e}

\usepackage{amsmath}
\usepackage{amsfonts}
\usepackage{amssymb}
\usepackage{theorem}
\usepackage{doublespace}
\usepackage{graphicx}
\usepackage{psfrag}
\usepackage{float}
\usepackage{subfigure}
\usepackage{mitfancyheadings}

\newcommand{\be}[1]{\begin{equation}\label{#1}}
\newcommand{\benon}{\begin{equation*}}  
\newcommand{\bemuln}[1]{\begin{multline}\label{#1}}
\newcommand{\bemul}{\begin{multline*}}
\newcommand{\bee}{\begin{eqnarray*}}
\newcommand{\eee}{\end{eqnarray*}}
\newcommand{\been}[1]{\begin{eqnarray}\label{#1}}
\newcommand{\eeen}{\end{eqnarray}}
\newcommand{\began}[1]{\begin{gather}\label{#1}}
\newcommand{\bega}{\begin{gather*}}
\newcommand{\bealn}[1]{\begin{align}\label{#1}}
\newcommand{\beal}{\begin{align*}}
\newcommand{\bealatn}[2]{\begin{alignat}{#1}\label{#2}}
\newcommand{\bealat}{\begin{alignat*}}
\newcommand{\bexalatn}[1]{\begin{xalignat}\label{#1}}
\newcommand{\bexalat}{\begin{xalignat*}}

\newcommand{\qed}{\newline \mbox{ } \hfill 
             \rule[-1pt]{2.5mm}{2.5mm}\par\vskip 10pt }
\newcommand{\ra}{\rightarrow}

\newcommand{\pf}{\vskip 5 pt \noindent {\bf Proof : }}

\newcommand{\mb}{\mathbf}

\newcommand{\mbb}{\mathbb}
\newcommand{\bosy}{\boldsymbol}

{\theoremstyle{plain} \newtheorem{thm}{Theorem}[section]

\newtheorem{prop}[thm]{Proposition}
\newtheorem{lemma}[thm]{Lemma}
}
{\theoremstyle{break} \theorembodyfont{\it}

\newtheorem{ass}{Assumption} }

\newlength{\captionwidth}
\newsavebox{\mytempbox}


\makeatletter   
\renewcommand{\subsubsection}{\@startsection 
{subsubsection}  
{3}                             
{-1em}                          
{-3.25ex plus -1ex minus-.2ex}  
{1.5ex plus.2ex}                
{\reset@font\normalsize\bf}}    
\renewcommand{\thesubsubsection}{}
\makeatother

\newcommand{\Ascr}{{\cal A}}
\newcommand{\Escr}{{\cal E}}

\newcommand{\Pscr}{{\cal P}}
\newcommand{\Iscr}{{\cal I}}

\makeatother

\begin{document}

\bibliographystyle{amsalpha} 

\title{Robust measurement-based buffer overflow probability estimators 
for QoS provisioning and traffic anomaly prediction applications}  

\author{\normalsize Spyridon Vassilaras \thanks{Spyridon Vassilaras. 
Athens Information Technology, 0.8km Markopoulo Ave., P.O. Box 68, Peania, 19002, Greece, e-mail: svas@ait.edu.gr
}
\normalsize and \ Ioannis Ch. Paschalidis 
\thanks{Ioannis Ch. Paschalidis, Dept. of Electrical \&
			Computer Eng., Division of Systems Eng., and Center for
			Information \& Systems Eng., Boston University, 8 St.
			Mary's St., Boston, MA 02215, e-mail:
                        yannisp@bu.edu, Web: http://ionia.bu.edu/. 
Research partially supported by the NSF under grants
EFRI-0735974, CNS-1239021, and 
	IIS-1237022, by the ARO under grants W911NF-11-1-0227 and
	W911NF-12-1-0390, and by the ONR under grant N00014-10-1-0952.
}}

\date{May 2012\\ \vspace{40pt}
}

\maketitle

\begin{abstract}
Suitable estimators for a class of Large Deviation approximations of rare event probabilities
based on sample realizations of random processes have been
proposed in our earlier work \cite{pava01}. 
These estimators are expressed as non-linear multi-dimensional optimization problems of a special structure.
In this paper, we develop an algorithm to solve these optimization problems very efficiently based on their characteristic structure. After discussing the nature of the objective function and constraint set and their peculiarities, we provide a formal proof that the developed algorithm is guaranteed to always converge.
The existence of efficient and provably convergent algorithms for solving these problems is a prerequisite for using the proposed estimators in real time problems such as call admission control, adaptive modulation and coding with QoS constraints, and traffic anomaly detection in high data rate communication networks.

\noindent {\bf Keywords: Large Deviations, Markov-modulated processes,
Estimation, Non-linear optimization, QoS, Traffic anomaly detection.} 
\end{abstract}

\thispagestyle{plain}

\section{Introduction} \label{sec:intro}

{\allowdisplaybreaks

{\em Large Deviations} (LD) theory (see \cite{deze2}) is an important analytical
tool that has been applied to provide approximate solutions to a variety 
of queuing and stochastic decision problems for which analytical solutions are
not practical (see for example \cite{bpt5,bpt67rev,pas4rev} and references
therein).
Most LD results provide an asymptotic approximation for the probability of a rare
event that is formulated as a non-linear optimization problem. Solving this optimization
problem is preferable to calculating the exact solution, which is often intractable or
requires computing a definite integral in a high dimensional space. This computation 
typically requires numerical integration since the integrant is not given in closed form.

In prior work \cite{pava01}, we have proposed a class of LD estimators of small overflow 
probabilities at a queue fed by Markovian arrival processes when the statistics of the
arrival process are estimated from a finite realization of the process. The proposed
estimators are expressed as non-linear multi-dimensional optimization problems of a special structure. In \cite{pava01}, a heuristic, custom-made algorithm to solve this type of problems was presented. Unfortunately, this algorithm, albeit working very well in most cases, did not exhibit guaranteed convergence. In this paper, we improve on the previous algorithm to develop an algorithm that maintains the efficiency of the previous one while guaranteeing convergence. Proving that this algorithm always converges to the global optimum is the main result of this paper.  
 
To better understand the context of this optimization problem, we start by a more
detailed description of its perceived applications. In modern high-speed fixed
communication networks congestion manifests itself as
buffer overflows; the {\em Quality of Service (QoS)} faced by various
connections can be quantified by the buffer overflow probability.  To provide
QoS guarantees the so called {\em effective bandwidth} admission control
mechanism has been proposed \cite{hui,gihu,kel2,guahna,kel3,pas4rev}. Briefly,
effective bandwidth is a number between the peak and average rate of a
connection such that when connections are allocated their effective bandwidth
in an appropriately dimensioned buffer, the buffer overflow probability stays
below a given small level (say on the order of $10^{-6}$).  Real-time
applications can tolerate such small frequencies of congestion phenomena. In
wireless networks, which experience higher Bit Error Rate (BER) and time varying
link quality, the targeted buffer overflow probability is much higher while the
effective bandwidth of the source cannot be offered to a connection on a
constant basis (since the available link capacity is a function of the link
quality). In order to characterize the time-varying behavior of a wireless link
the dual concept of {\em effective capacity} has been proposed by \cite{wunegi1,wu}. The effective capacity of a wireless link is the maximum constant source rate
that can be offered to this link such that the buffer overflow probability stays
below a given level. As the data rate and BER over a wireless link depend on the
modulation and coding scheme applied, adaptive modulation and coding can achieve
maximum effective capacity under a given BER constraint
\cite{liu_giann1,liu_giann2,liu_giann3,tang_zhang1,tang_zhang2,tang_zhang3}.
More recently, a combined buffer overflow probability and BER minimization scheme
has been proposed in \cite{svas10}.      

In make-to-stock manufacturing systems and supply chains, 
the objective is to control the inventory in order to avoid stockouts (see
\cite{bepa1,gla97,pali1}). In such systems demand is met from a finished goods
inventory, and it is backlogged if inventory is not available. The stockout
probability quantifies the QoS encountered by customers. It can be shown that
this probability is equal to a buffer overflow probability in a corresponding
make-to-order system \cite{bepa1,pali1}. Thus, the problem of estimating the
stockout probability can be transformed into one of estimating a buffer
overflow probability.

Similar LD techniques have been applied in traffic anomaly
detection where real time monitoring and analysis of the
aggregate traffic in various points of a network can identify and
classify unusual traffic fluctuations that can be caused by
network intrusion, malicious attacks, network malfunction or
irregular use of network resources
\cite{pas-sma-ton-09,pas-chen-tosn-10}. LD techniques for traffic
anomaly detection can provide automatic alerts to previously
unknown suspicious events, unlike signature based techniques that
rely on known malicious content in data packets.

In each one of the above applications, we are interested in estimating buffer overflow or
excessive queuing delay or stockout or traffic anomaly indicator probabilities that are very small. Moreover, arrival and service processes are typically autocorrelated (to model bursty traffic in communication networks and time varying capacity of wireless links, and to accommodate realistic demand scenarios and model failure-prone production facilities in make-to-stock manufacturing systems). As a result,
obtaining exact analytic expressions is intractable; it is, therefore, natural
to focus on approximate solutions such as the ones developed using LD
techniques.
 
Most of the initial LD work assumed the detailed knowledge of models for
the arrival and service processes.  In practice however, such traffic models
are not known a priori and have to be estimated from real
observations. Consequently, one approach for estimating buffer overflow
probabilities is to assume a certain traffic model, estimate the parameters of the 
model from real observations, and then calculate the overflow probability using large
deviations results. Consider for example the case of a {\em deterministic
Markov-modulated} source (i.e., a source which has a constant rate at every
state of the underlying Markov chain) and let $g(\bosy{\Xi})$ be the overflow
probability when this source is fed in a certain buffer, where $\bosy{\Xi}$
denotes the transition probability matrix of the underlying Markov-chain. (We
assume that the rate at every state, and the characteristics of the buffer are
given, thus,  we do not explicitly denote the dependence of the overflow
probability on these quantities). Let $\hat{\bosy{\Xi}}$ be an unbiased
estimator of the transition probability matrix $\bosy{\Xi}$, that is,
${\mathbf E}[\hat{\bosy{\Xi}}] = \bosy{\Xi}$. Suppose that we use
$g(\hat{\bosy{\Xi}})$ as an estimator for the overflow probability. An
important observation is that due to the nonlinearity of $g(\cdot)$, ${\mathbf
E}[g(\hat{\bosy{\Xi}})]$ is not necessarily equal to $g({\mathbf
E}[\hat{\bosy{\Xi}}])=g(\bosy{\Xi})$. That is, a {\em certainty equivalence}
approach can lead to an erroneous estimate. 

Subsequent work in the field has recognized this practical issue and developed 
LD estimators for measurement-based, model-free admission control \cite{grts1,grts2,duff1}.
Our prior work \cite{pava01} proposed new estimators of overflow probabilities in queues 
fed by Markov-modulated arrival processes. Intuitively, the proposed estimators 
suggested that we should quote a quantity that is larger than $g(\hat{\bosy{\Xi}})$ to guard
against estimation errors. Hence, these estimators are ``safe'' even when based on relatively few
observations, meaning that they do not lead to substantial underestimation of
the overflow probability that can compromise QoS.  Still, they are {\em
consistent} in the sense that they converge to $g(\bosy{\Xi})$ with
probability 1 (w.p.1) as the number of observations tends to infinity.

The development of efficient non-linear optimization algorithms to compute the proposed
estimators is of paramount importance. Given that these estimators are meant to be used
for real time applications in fixed and wireless telecommunication networks,
they should be computed in very short time. This means that the non-linear optimization algorithm must guarantee convergence to the minimum in a small number of steps. The high dimensionality of the decision variable space and the nature of the objective function (which is computationally expensive to calculate) makes standard optimization algorithms not efficient enough. The algorithm proposed in this paper takes advantage of the special structure of the objective
function and the constraint set to achieve fast and guaranteed convergence.   

This paper is organized as follows. We start in Section~\ref{sec:form} with
a summary of the key results and notation from \cite{pava01} including the optimization problems that we want to solve. 
In Section~\ref{sec:NLP} we analyze the structure of the objective functions and constraint sets of these optimization problems and discuss their key properties such as differentiability and convexity.  In Section~\ref{sec:algo} we develop two algorithms for solving the optimization problems and show that the second one (which is an improved version of the first one) exhibits guaranteed convergence. Finally, conclusions are in Section~\ref{sec:concl}.

On a notational remark, in this paper, matrices are denoted by bold uppercase 
and vectors by bold lowercase characters. Moreover, all vectors will be 
assumed to be column vectors, unless otherwise specified. 

\section{Estimates of the overflow probability based on an arrival process realization}
\label{sec:form}

In this Section we summarize the key results and notation from \cite{pava01} in order to establish our notation and understand the significance of the optimization problems at hand. Consider a single class G/G/1 queue with a Markov-modulated arrival process. We will be using a discrete-time model, where
time is divided into time slots. We let $A_k$, $k\in \mbb{Z}$, denote the
aggregate number of ``customers'' that enter the queue at time $k$. The queue has
an infinite buffer and is serviced according to a general service process
which can clear up to $B_k$ customers during the time interval $[k,k+1]$.  We
assume that the stochastic processes $\{A_k;\ k\in \mbb{Z}\}$ and $\{B_k;\
k\in \mbb{Z}\}$ are stationary, possibly
autocorrelated and mutually independent processes. For such a discrete time 
stochastic process $\{X_k;\ k\in \mbb{Z}\}$ let us define the partial sum process 
$S_n=\sum_{k=1}^n X_k$. We assume that the partial sum processes of both the 
arrival and service processes  $\{A_k;\ k\in \mbb{Z}\}$ and $\{B_k;\
k\in \mbb{Z}\}$ satisfy some technical conditions (see Assumption A in \cite{pava01}) among which is the property that the limit 
$$\Lambda_X(\theta)\stackrel{\triangle}{=} 
\lim_{n\rightarrow \infty}  \Lambda_n^X(\theta) =
\lim_{n\rightarrow \infty} \frac{1}{n} 
\log {\mathbf E}[e^{\theta S_n}],$$  
exists for all $\theta$, where $\pm \infty$ are allowed both
as elements of the sequence $\Lambda_n^X(\theta)$ and as limit points. Note that the set of processes satisfying these conditions  is large enough to include renewal, Markov-modulated,
and stationary processes with mild mixing conditions. Such processes can model
``burstiness'' and are commonly used in modeling the input traffic to
communication networks and the link capacity fluctuations due to wireless channel fading. 
They have also being used in modeling demand and
the production process in manufacturing systems \cite{bepa1,pali1}.

Let us define:
\be{L*} \Lambda_X^*(a)\stackrel{\triangle}{=} 
        \sup_{\theta}(\theta a-\Lambda_X(\theta)), \end{equation}
which is the Legendre transform of $\Lambda_X(\cdot)$.  
The function $\Lambda_X^*(\cdot)$ is convex and lower semicontinuous (see
\cite{deze2}). In Large Deviations parlance,
$\Lambda_{X}(\cdot)$ and $\Lambda^*_{X}(\cdot)$ are called the limiting
$\log$-moment generating function and the large deviations rate function,
respectively, of the process $\{X_k;\ k\in \mbb{Z}\}$.  

We denote by $L_k$ the queue length at time $k$ (without counting arrivals at
time $k$). We assume that the server uses a work-conserving policy (i.e., the
server never stays idle when there is work in the system) and that 
\be{stable} {\mathbf E}[A_1] < {\mathbf E}[B_1], \end{equation} 
which by stationarity carries over to all $k$.  We also assume that the queue
length process $\{L_k; k\in\mbb{Z}\}$ is stationary.  To simplify the analysis
we consider a discrete-time ``fluid'' model, meaning that we will be treating
$A_k$, $L_k$, and $B_k$ as nonnegative real numbers (the amount of fluid
entering the buffer, in queue, and the service capacity, respectively).

A Large Deviations Principle (LDP) for the queue length process has been established in 
\cite{glwh,bpt5} and is given in the next proposition.  In preparation for the result, 
consider a convex function $g(u)$ with the property $g(0)=0$. We define the {\em
largest root} of $g(u)$ to be the solution of the optimization problem
$\sup_{u: g(u)< 0} u$.  If $g(\cdot)$ has negative derivative at $u=0$, there
are two cases: either $g(\cdot)$ has a single positive root or it stays below
the horizontal axis for all $u>0$. In the latter case, we will say that
$g(\cdot)$ has a root at $u=\infty$.

\begin{prop} \label{prop:ldp-queue} 
The steady-state queue length process $L_k$ satisfies
\be{tail-1cl} \lim_{U\ra \infty} \frac{1}{U} \log {\mathbf P}[L_k\geq U] =
-\theta^*, \end{equation}
where $\theta^*>0$ is the largest root of the equation
\be{root} \Lambda_A(\theta)+\Lambda_B(-\theta)=0. \end{equation}
\end{prop}
More intuitively, for large enough $U$ we have 
$$ {\mathbf P}[L_k\geq U] \sim e^{-U \theta^*}. $$ 
This expression can be used to estimate the overflow probability in a queue
with a finite buffer of size $U$. Kelly \cite{kel3} establishes that the
latter probability has the same asymptotic decay rate (same exponent) with
${\mathbf P}[L_k\geq U]$.

\subsection{Markov-modulated arrivals}

Calculating the limiting log-moment generating function can be made easier than computing 
infinite sums when dealing with certain special stochastic processes.
The most common such process is the Markov-modulated process, which is defined as follows: 
Consider an irreducible Markov chain with $M$
states $1,2,\ldots,M$ and transition probability matrix $\bosy{\Xi}
= \linebreak[3] \{ p(i,j)\}_{i,j=1}^M$. 
We will be using the notation ${\mb{p}}^T=({\mb{p}}_{1}^T,\ldots,{\mb{p}}_{M}^T)$, 
where ${\mb{p}}_i^T$ is the $i$-th row of $\bosy{\Xi}$ and $\bosy{A}^T$ denotes the transpose of $\bosy{A}$.
The Markov chain makes one transition per time slot; let $Y_k$
be the state at time $k$. The number of arrivals at time $k$ is a
random function of the state, i.e., it is drawn according to a p.d.f. 
$f_{Y_{k}}(\cdot)$ associated with that state. Let us denote by
$\eta_i(\theta)$ the moment generating function of $f_i(\cdot)$. 
Note that for a deterministic amount of arrivals $r_i$ per time slot at state $i$, the moment generating function is $\eta_i(\theta) = e^{\theta r_i}$.

In \cite[Sec. 3.1.1]{deze2} it is established that the limiting $\log$-moment
generating function of the arrival process $\{A_k;\ k\in \mbb{Z}\}$, is given by 
\be{lambdaA} \Lambda_A(\theta,{\mb{p}}) = \log
\rho(\bosy{\Pi}^A_{\theta,{\mb{p}}}), 
\end{equation} 
where $\rho(\bosy{\Pi}^A_{\theta,{\mb{p}}})$ denotes the Perron-Frobenius
eigenvalue of the $M\times M$ matrix: 
\be{MMLDP}
\bosy{\Pi}^A_{\theta,{\mb{p}}} =
\{p(i,j) \eta_j(\theta)\}_{i,j=1}^M.
\end{equation} 
(In this Markov-modulated case we are using notation that explicitly denotes
the dependence of $\Lambda_A(\theta,{\mb{p}})$ and
$\bosy{\Pi}^A_{\theta,{\mb{p}}}$ 
on the transition probabilities ${\mb{p}}$.)
Notice that because the quantities $\eta_j(\theta)$ are always positive 
the irreducibility of $\bosy{\Xi}$ implies that
$\bosy{\Pi}^A_{\theta,{\mb{p}}}$ is irreducible.

\subsection{Estimating the Overflow Probability}

Consider now the more realistic case where the transition probabilities $p(i,j)$
are not known in advance, but need to be estimated by observing the arrival process.
In particular, we will be assuming that we have perfect
knowledge of the service process $\{B_k;\ k\in \mbb{Z}\}$, and that we can observe the states of
the Markov chain associated with the arrival process. That is, we do know $M$
and the probability density functions $f_i(\cdot)$ (the supports of whom do not overlap), 
but the transition probability matrix
$\bosy{\Xi}$ is unknown. Suppose that we observe a sequence 
${\mb{Y}}=Y_1,Y_2,\ldots,Y_n$ of states that the unknown
Markov chain visits with the initial state being $Y_0=\sigma$.
Consider the empirical measures:
$$ q_{n}^{\mb{Y}} ({\mb{y}}) = \frac{1}{n} \sum_{k=1}^n
{\mb{1}}_{\mb{y}}(Y_{k-1}Y_k),$$  
where ${\mb{y}}\in \Ascr^2 \stackrel{\triangle}{=} \{1,\ldots,M\}\times
\{1,\ldots,M\}$. Note that when ${\mb{y}}=(i,j)\in \Ascr^2$ the empirical
measure $q_{n}^{\mb{Y}} ({\mb{y}})$ denotes the fraction of times that the
Markov chain makes transitions from $i$ to $j$ in the sequence
${\mb{Y}}$. Let now $\Ascr^2_{\mb{p}}\stackrel{\triangle}{=}\{(i,j)\in \Ascr^2\mid
p(i,j)>0\}$ denote the set of pairs of states that can appear in the sequence
$Y_1,Y_2,\ldots,Y_n$ and denote by $M_1(\Ascr^2_{\mb{p}})$ the standard
$|\Ascr^2_{\mb{p}}|$-dimensional probability simplex, where $|\Ascr^2_{\mb{p}}|$
denotes the cardinality of $\Ascr^2_{\mb{p}}$. 
Note that the vector  
of $q_{n}^{\mb{Y}}({\mb{y}})$'s denoted by $\bosy{q}_{n}^{\mb{Y}} =
(q_{n}^{\mb{Y}}({\mb{y}});\ {\mb{y}}\in \Ascr^2_{\mb{p}})$ is an element of
$M_1(\Ascr^2_{\mb{p}})$. For any ${\mb{q}} \in M_1(\Ascr^2_{\mb{p}})$, let  
\be{q-MM} q_1(i) \stackrel{\triangle}{=} \sum_{j=1}^M q(i,j) \qquad \text{and}
\qquad q_2(i) = \sum_{j=1}^M q(j,i) 
\end{equation}
be its marginals. Whenever $q_1(i)>0$, let $q_f(j\mid i)
\stackrel{\triangle}{=} q(i,j)/q_1(i)$. We will be using the notation 
${\mb{q}}_f = (q_f(1\mid 1), \ldots, q_f(M\mid 1), q_f(1\mid 2),\ldots,
q_f(M\mid 2), \ldots, q_f(1\mid M),\ldots, q_f(M\mid M))$. (To avoid overburdening the notation, we will 
suppress the dependence of the estimators on the sequence ${\mb{Y}}$. We will also often ommit the subscript $n$ and simply write  ${\mb{q}}$, ${\mb{q}}_{f}$ and $q_{1}(i)$.)

Note that ${\mb{p}}$ (the vector of the actual but unknown transition probabilities) is an element of $(M_1(\Ascr))^M$ (i.e.,
the $M$-times cartesian product of $M_1(\Ascr)$). As the
transition probabilities ${\mb{p}}$ are not known, we assume a prior pdf
$\phi_{{\mb{p}}^-}\in (M_1(M_1(\Ascr)))^M$, which assigns probability mass only
to ${\mb{p}}$'s corresponding to irreducible Markov chains. Let
$\underline{\mb{p}}^-$ denote the support of $\phi_{{\mb{p}}^-}$. 

Let us define:

\be{eq:thm-est-rate} 
I_3({\mb{p}}) = \begin{cases} \sum_{i=1}^M q_{1}(i) H( q_{f}(\cdot\mid i) \mid
p(i,\cdot) ), & \text{if ${\mb{p}}\in \underline{\mb{p}}^-$,} \\   
\infty, & \text{otherwise.} \end{cases}
\end{equation}

where $H( q_f(\cdot\mid i) \mid p(i,\cdot) )$ is the relative entropy defined
as 
$$ H( q_f(\cdot\mid i) \mid p(i,\cdot) ) = \sum_{j=1}^M q_f(j\mid i) \log
\frac{q_f(j\mid i)}{p(i,j)}. $$  

Now given the empirical measure ${\mb{q}}_n$, a maximum likelihood
estimator of the transition probabilities is given by  
\be{mle} \hat{p}_n(i,j) = q_{nf}(j\mid i) = \frac{q_n(i,j)}{q_{n1}(i)},
\qquad i,j=1,\ldots,M,
\end{equation}
where the extra $n$ in the subscripts explicitly denotes the dependence on the length $n$ of the sample sequence.  
Let $\hat{\mb{p}}_n$ denote the vector of these estimates. 
We can now construct a
matrix $\bosy{\Pi}^A_{\theta,\hat{\mb{p}}_n}$ with elements
$$ \pi^A_{\theta,\hat{\mb{p}}_n}(i,j) = \hat{p}_n(i,j) \eta_j(\theta)\,
\qquad i,j=1,\ldots,M, $$
and obtain an estimate,
$\Lambda_{A}(\theta,\hat{\mb{p}}_n)$, of the limiting $\log$-moment
generating function 
for the arrival process by computing the Perron-Frobenius eigenvalue of
$\bosy{\Pi}^A_{\theta,\hat{\mb{p}}_n}$.
The following 3 estimators of the overflow probability ${\mathbf P}[L_i\geq U]$ have been described in \cite{pava01} (the reader is referred to \cite{pava01} for more detailed explanation of  these estimators).

The traditional ``certainty equivalent" estimator:
\be{loss-e1} \Pscr^I_n \stackrel{\triangle}{=}
e^{-U\theta^*(\hat{\mb{p}}_n)}, 
\end{equation} 
where $\theta^*(\hat{\mb{p}}_n)$ is the largest root of the equation
$\Lambda_{A}(\theta,\hat{\mb{p}}_n)+\Lambda_B(-\theta)=0$.  

And two improved estimators:
\be{loss-e2} \Pscr^{II}_n \stackrel{\triangle}{=}
\exp\biggl\{-n\inf_{{\mb{p}}\in (M_1(\Ascr))^M} [ s\theta^*({\mb{p}})+
I_3({\mb{p}})]\biggr\}. 
\end{equation}

and

\be{loss-e4} \Pscr^{IV}_n(\mu) \stackrel{\triangle}{=} \Pscr^{II}_n + \mu \sqrt{\Pscr^{III}_n(2) -
(\Pscr^{II}_n)^2}, 
\end{equation}
where $\mu$ is some scalar and

\be{loss-e3} \Pscr^{III}_n(\ell) \stackrel{\triangle}{=}  
\exp\biggl\{-n\inf_{{\mb{p}}\in (M_1(\Ascr))^M} [ \ell s\theta^*({\mb{p}})+
I_3({\mb{p}})]\biggr\} \qquad \ell =1,2,\ldots. 
\end{equation}

 $\Pscr^{IV}_n(\mu)$ can be interpreted as expectation plus $\mu$ times standard
deviation. The idea is that by selecting $\mu$ large enough $\Pscr^{IV}_n(\mu)$ can be adequately safe, 
i.e., the likelihood that $\Pscr^{IV}_n(\mu)$ underestimates the true probability of loss can be made 
adequately small.

\section{Computing the Estimators} \label{sec:NLP}

Computing the estimators in (\ref{loss-e2}) and (\ref{loss-e3}) requires
solving nonlinear optimization problems. In this section we examine the
structure of these problems and devise efficient algorithms for their
solution. 

The optimization problems in (\ref{loss-e2}) and (\ref{loss-e3}) have the
following form 
\begin{equation} \label{NLP1}
\begin{array}{rl}
\text{minimize} & \ell s\theta^{*}({\mb{p}})+I_3({\mathbf{p}}) \\ 
\text{s.t.} &p(i,j)\geq 0, \qquad \quad i,j=1,\ldots,M, \\ 
& \sum_{j=1}^{M} p(i,j)=1, \quad \ i=1,\ldots,M,
\end{array}
\end{equation}
where $\ell$ is some positive scalar and $I_3({\mb{p}})$ is given by 
(cf. Eq.~(\ref{eq:thm-est-rate})) 
\be{eq:est-rate} 
I_3({\mb{p}}) = \begin{cases} \sum_{i=1}^M q_{1}(i) \sum_{j=1}^M q_f(j\mid i)
\log\frac{q_f(j\mid i)}{p(i,j)}, \qquad \qquad 
\text{if ${\mb{p}}\in \underline{\mb{p}}^-$,} \\    
\infty, \qquad \qquad \qquad \qquad \qquad \qquad \qquad \qquad \qquad  \text{otherwise.} \end{cases}
\end{equation}

We are interested in solving relatively large instances of (\ref{NLP1});
typically $M$ can be in the range of 5-100 which brings the number of
decision variables in the range of 25-10000. Since one major application for these 
estimators is Call Admission Control in telecommunication networks that provide QoS guarantees,
the optimization problems at hand should be solved in real time. 
As a result, computational
efficiency is critical and special purpose algorithms that exploit the special
structure of these optimization problems are of interest.

On the structure of the objective function, a first observation is that
$I_3(\mb{p})$ is a convex function if $\underline{\mb{p}}^-$ is a convex
set. This can be easily established directly from the definition of
convexity. Furthermore, $I_3({\mb{p}})$ is strictly convex in
$\underline{\mb{p}}^-$ (where it is finite). In Section~\ref{sec:form} we
have assumed that the prior assigns probability mass only to ${\mb{p}}$'s
corresponding to irreducible Markov chains. Let $\Iscr$ be the set of
${\mb{p}}$'s corresponding to irreducible Markov chains. 
The set $\Iscr$ is a convex set (for a proof see \cite{pava01}). 
  
Based on the convexity of $\Iscr$ and assuming that the support of the prior
$\underline{\mb{p}}^-$ is a convex subset of $\Iscr$, $I_3({\mb{p}})$ is a
convex function. Henceforth, and in the absence of more information on the true
transition probabilities of the Markov chain we wish to estimate, we will be
making the following assumption.
\begin{ass}\label{supp-I}
The support of the prior, $\underline{\mb{p}}^-$, is the set $\Iscr$ of
transition probability vectors ${\mb{p}}$ corresponding to irreducible Markov
chains.  
\end{ass}

Assumption \ref{supp-I} implies that $n$ can always be taken large enough so that 
at least one exit transition for each state has been observed and therefore
${\mb{q}}_f$ is the transition probability vector of an irreducible Markov chain.  
Hence we assume that if the observed sequence ${\mb{Y}}$ results to 
${\mb{q}}_f \notin \Iscr$ then we prolong the observation period by a few samples to
achieve ${\mb{q}}_f \in \Iscr$. 

Consider now the following optimization problem:
\begin{equation} \label{NLP1-m}
\begin{array}{rl}
\text{minimize} & \ell s\theta^{*}({\mb{p}})+\hat{I}_3({\mathbf{p}}) \\ 
\text{s.t.} &p(i,j)\geq 0, \qquad i,j=1,\ldots,M, \\ 
& \sum_{j=1}^{M} p(i,j)=1, \qquad i=1,\ldots,M,
\end{array}
\end{equation}
where 
\be{eq:est-rate-m} 
\hat{I}_3({\mb{p}}) = \sum_{i=1}^M q_{1}(i) \sum_{j=1}^M q_f(j\mid i)
\log\frac{q_f(j\mid i)}{p(i,j)}.
\end{equation}

The following property of the optimal solution of the above optimization problem is established in \cite{pava01}:
 
\begin{lemma}
Let ${\mb{p}}^*$ be an optimal solution of the optimization problem in
(\ref{NLP1-m}). Then ${\mb{p}}^*$ is the transition probability vector of an
irreducible Markov chain. 
\end{lemma}

The result of this lemma suggests that under Assumption~\ref{supp-I} the
optimization problem in (\ref{NLP1-m}) is equivalent to the problem in
(\ref{NLP1}). Hence, we will focus on solving (\ref{NLP1-m}).

A very important issue for any nonlinear programming problem is the form of
the objective function. A convex objective function, especially under
polyhedral constraints (as in (\ref{NLP1-m})), can lead to more efficient
algorithms. Let 
$$\bar{\theta}({\mb{p}})\stackrel{\triangle}{=}
\ell s\theta^{*}({\mb{p}})+\hat{I}_3({\mathbf{p}})$$
denote the objective function in (\ref{NLP1-m}). 
Notice that it is the weighted sum of a strictly convex
function,  $\hat{I}_3({\mathbf{p}})$, and $\theta^{*}({\mathbf{p}})$, which is not
necessarily convex. To see this, we consider an example where $\ell =1$ and the arrival 
process is a two-state Markov-modulated process with deterministic 
number of arrivals per state
$(r_1, r_2)=(0.042,0.077)$. This arrival process is fed into a buffer which is served at fixed 
service rate $c=0.058$. The parameter $s$ was set equal to $0.002$. In 
Figure~\ref{fig:th*-noncnvx} (a) we
plot $\theta^{*}({\mathbf{p}})$ and in (b) $\bar{\theta}({\mathbf{p}})$ versus the
two (arbitrarily chosen as) independent decision variables $p(1,1)$ and
$p(2,2)$. It can be seen that although $\theta^{*}({\mathbf{p}})$ is not convex
(it is convex only along some directions), $\bar{\theta}({\mathbf{p}})$ is
convex. Consequently, $\bar{\theta}({\mb{p}})$
is not convex in general. Nevertheless, since  $\hat{I}_3({\mathbf{p}})$ is
strictly convex, it can be seen that $\bar{\theta}({\mb{p}})$ will be convex for
small enough values of $s$. Recalling that $s=U/n$ and assuming that the
buffer size $U$ is given, we will be dealing with a convex objective function
if we can afford a large number $n$ of measurements. 
\begin{figure}[ht]
\begin{center}
\mbox{\subfigure[\mbox{}]{
\psfrag{p(1,1)}{\small $p(1,1)$}
\psfrag{p(2,2)}{\small $p(2,2)$}
\psfrag{thStar(p)}{\small $\theta^{*}({\mathbf{p}})$}
\includegraphics[width=0.45\textwidth,totalheight=0.35\textwidth]{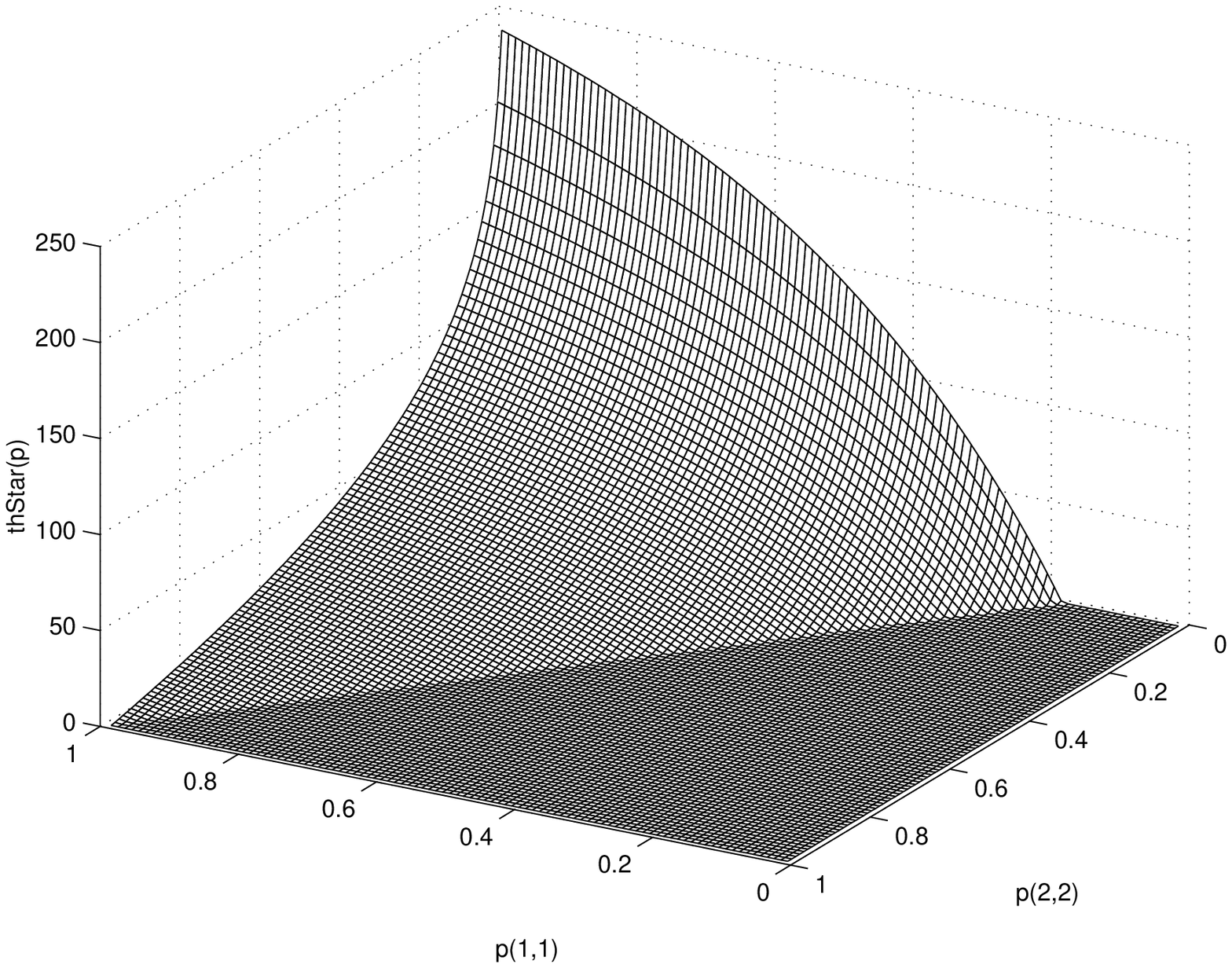}} 
\subfigure[\mbox{}]{
\psfrag{p(1,1)}{\small $p(1,1)$}
\psfrag{p(2,2)}{\small $p(2,2)$}
\psfrag{s*thStar(p) + I3(p)}{\small $\bar{\theta}({\mathbf{p}})$}
\includegraphics[width=0.45\textwidth,totalheight=0.35\textwidth]{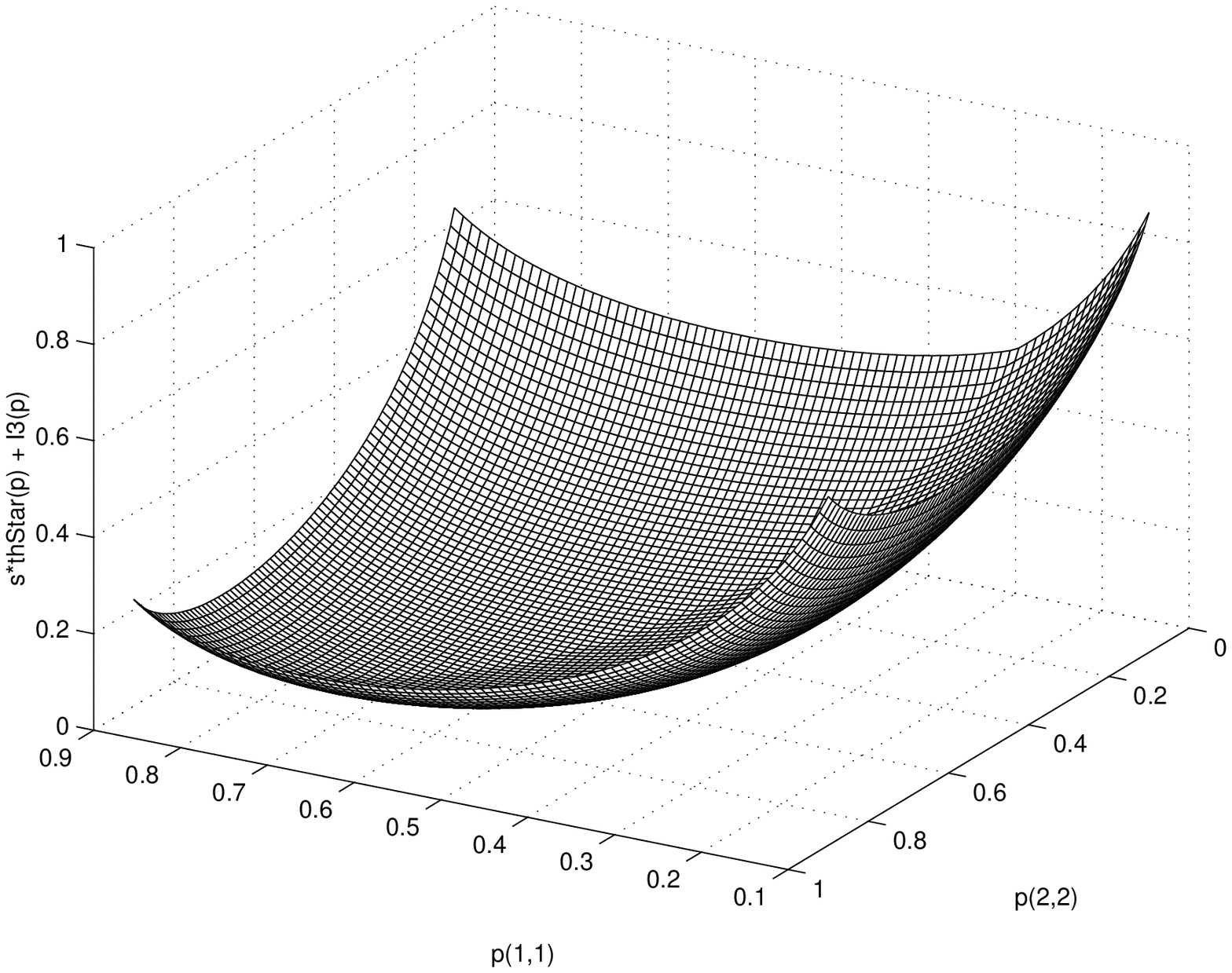}}}
\end{center}
\caption{Plots of $\theta^{*}({\mathbf{p}})$ (a) and $\bar{\theta}({\mathbf{p}})$ (b) 
versus $p(1,1)$ and $p(2,2)$ when the arrival process is a Markov chain with 2 states.}
\label{fig:th*-noncnvx}
\end{figure}

Another important observation, which will affect the convergence
of the optimization algorithms presented in the next section, is that 
$\theta ^{*}({\mathbf{p}})$ might be equal to $0$ (which translates into 
${\mathcal{P}}_{n}^{I}=1$) for some values of ${\mathbf{p}}$. This happens when
the mean arrival rate is larger or equal to the mean service rate ${\mathbf{%
E}}[B_1]$ (which is equal to $c$ in the most common case of a deterministic
service process). Therefore, if we denote by $p_{1}(i)$ the steady state
probability for state $i$ corresponding to the transition probability vector 
${\mathbf{p}}$ and by ${\mathbf{E}}[A_1 \mid i]$ the conditional mean number of
arrivals given that we are at state $i$, the condition for $\theta ^{*}(%
{\mathbf{p}})=0$ can be written as:

\begin{equation}
\sum_{i=1}^{M}p_{1}(i)\text{ }{\mathbf{E}}[A_1 \mid i]\geq {\mathbf{E}}[B_1]
\label{zerothstar1}
\end{equation}

Let us now define the following subsets of ${\mathcal{I}}$:

\begin{itemize}
\item []  
\item   ${\mathcal{I}}_{1} \triangleq \{{\mathbf{p}}\in {\mathcal{I}} \mid
\sum_{i=1}^{M}p_{1}(i)$ ${\mathbf{E}}[A_1 \mid i]<{\mathbf{E}}[B_1]\}$
\item []  
\item  ${\mathcal{I}}_{2}\triangleq \{{\mathbf{p}}\in {\mathcal{I}} \mid
\sum_{i=1}^{M}p_{1}(i)$ ${\mathbf{E}}[A_1 \mid i]>{\mathbf{E}}[B_1]\}$
\item []  
\item  ${\mathcal{I}}_{3}\triangleq \{{\mathbf{p}}\in {\mathcal{I}} \mid
\sum_{i=1}^{M}p_{1}(i)$ ${\mathbf{E}}[A_1 \mid i]={\mathbf{E}}[B_1]\}$
\item []  
\item  ${\mathcal{I}}_{4}\triangleq \{{\mathbf{p}}\in {\mathcal{I}} \mid
\sum_{i=1}^{M}p_{1}(i)$ ${\mathbf{E}}[A_1 \mid i]\leq {\mathbf{E}}[B_1]\}$ $={\mathcal{I%
}}_{1}\cup {\mathcal{I}}_{3}$
\item []  
\item  ${\mathcal{I}}_{5}\triangleq \{{\mathbf{p}}\in {\mathcal{I}} \mid
\sum_{i=1}^{M}p_{1}(i)$ ${\mathbf{E}}[A_1 \mid i]\geq {\mathbf{E}}[B_1]\}$ $={\mathcal{I%
}}_{2}\cup {\mathcal{I}}_{3}$
\item []  
\end{itemize}

Obviously, ${\mathcal{I}}_{1}$, $ {\mathcal{I}}_{2}$ and ${\mathcal{I}}_{3}$ form a
partition of ${\mathcal{I}}$ into disjoint sets. Furthermore, the dimension
of ${\mathcal{I}}_{3}$ is always equal to the dimension of ${\mathcal{I}}$
minus one. ${\mathcal{I}}_{3}$ forms a boundary between ${\mathcal{I}}_{1}$ and $%
{\mathcal{I}}_{2}$. Note that if ${\mathbf{q}}_{f}\in {\mathcal{I}}_{5}$, then $%
{\mathcal{P}}_{n}^{I}=1$, which implies that ${\mathcal{P}}_{n}^{II}=1$ (since 
${\mathcal{P}}_{n}^{II}\geq {\mathcal{P}}_{n}^{I}$), which is a case of hardly any interest.
In the sequel, we will assume that ${\mathbf{q}}_{f}\in {\mathcal{I}}_{1}$.
The next lemma reveals an important property of ${\mathcal{I}}_{1}$ with
respect to the optimization problem in (\ref{NLP1-m}):

\begin{lemma} If ${\mathbf{q}}_{f}$ $\in {\mathcal{I}}_{1}$ then ${\mathbf{p}}%
^{*}\in {\mathcal{I}}_{4}$, where ${\mathbf{p}}^{*}$ is the minimizing ${\mathbf{p%
}}$ in (\ref{NLP1-m}). \label{sameSet}
\end{lemma}

\pf Assume ${\mathbf{p}}^{*}\notin {\mathcal{I}}_{4}$, which is equivalent to ${\mathbf{p}}%
^{*}\in {\mathcal{I}}_{2}$. Now, denote by $l_{1}$ the line segment that
connects the two points ${\mathbf{q}}_{f}$ and ${\mathbf{p}}^{*}$. For every
point ${\mathbf{p}}$ $\in l_{1}$ different than ${\mathbf{p}}^{*}$, $\hat{I}_{3}(%
{\mathbf{p}})<\hat{I}_{3}({\mathbf{p}}^{*})$ due to the strict convexity of $%
\hat{I}_{3}(\cdot )$ whose global minimum is at ${\mathbf{q}}_{f}$. Also, for every point ${\mathbf{p}}$ $\in l_{1}\cap $ $%
{\mathcal{I}}_{5}$, $\theta ^{*}({\mathbf{p}})=0$. Furthermore, there is at
least one point ${\mathbf{p}}$ $\in l_{1}\cap $ ${\mathcal{I}}_{5} \setminus \{{\mathbf{p}}%
^{*}\}$, since ${\mathbf{p}}^{*}\in {\mathcal{I}}_{2}$. For this point we have
that $\bar{\theta}({\mathbf{p}})=ks\theta ^{*}({\mathbf{p}})+$ $\hat{I}_{3}(%
{\mathbf{p}})=\hat{I}_{3}({\mathbf{p}})<\hat{I}_{3}({\mathbf{p}}^{*})=\bar{\theta}(%
{\mathbf{p}}^{*})$. But this contradicts our initial assumption that ${\mathbf{p}}^{*}$
is the minimizer in (\ref{NLP1-m}). Thus, ${\mathbf{p}}^{*}\in {\mathcal{I}}_{4}$.
\qed

Figure~\ref{fig:th*-noncnvx} seems to suggest that ${\mathcal{I}}_{3}$ is a
straight line, which makes ${\mathcal{I}}_{1}$, ${\mathcal{I}}_{2},{\mathcal{I}}_{4}
$ and ${\mathcal{I}}_{5}$ convex sets. Indeed, for the case where $M=2$, $%
p_{1}(1)$ and $p_{1}(2)$ can be expressed as a function of $p(1,1)$ and $%
p(2,2)$ as follows:
$$
p_{1}(1)=\frac{1-p(2,2)}{2-p(1,1)-p(2,2)} 
$$
and
$$
p_{1}(2)=\frac{1-p(1,1)}{2-p(1,1)-p(2,2)} 
$$
Therefore the condition for ${\mathbf{p}}\in {\mathcal{I}}_{3}$ becomes:
$$
\sum_{i=1}^{2}\text{ }\frac{1-p(i,i)}{2-p(1,1)-p(2,2)}{\mathbf{E}}[A_1 \mid 3-i]=%
{\mathbf{E}}[B_1]
$$
which after some routine algebraic manipulations can be written as:
$$
({\mathbf{E}}[B_1]-{\mathbf{E}}[A_1 \mid 2])p(1,1)+({\mathbf{E}}[B_1]-{\mathbf{E}}%
[A_1 \mid 1])p(2,2)=2{\mathbf{E}}[B_1]-{\mathbf{E}}[A_1 \mid 1]-{\mathbf{E}}[A_1 \mid 2]
$$
The above equation is clearly a line equation on the plane and that is in agreement
with the straight line border between ${\mathcal{I}}_{1}$ and ${\mathcal{I}}_{2}$
shown in Figure~\ref{fig:th*-noncnvx}. However, for $M>2$, it can be easily
shown by example that both sets ${\mathcal{I}}_{1}$ and ${\mathcal{I}}_{2}$ are
not necessarily convex. For a case with $M$=3, where ${\mathcal{I}}_{1}$ is not 
convex, take the following two elements of ${\mathcal{I}}$ given in matrix form \\
\\

\qquad $ {\bosy \Xi _1} =
\begin{bmatrix}
0.2 & 0.6 & 0.2 \\ 
0.6 & 0.2 & 0.2 \\ 
0.2 & 0.4 & 0.4 
\end{bmatrix} $
\qquad and \qquad
$ {\bosy \Xi _2} =
\begin{bmatrix}
0.7 & 0.1 & 0.2 \\ 
0.2 & 0.1 & 0.7 \\ 
0.3 & 0.6 & 0.1 
\end{bmatrix}$, \\

and assume 
$$[{\mathbf{E}}[A_1 \mid 1], [{\mathbf{E}}[A_1 \mid 2], [{\mathbf{E}}[A_1 \mid 3]]=
[0.2,1.0,2.5].$$
Figure~\ref{fig:noncon}(a) shows the mean arrival rate of a Markov-modulated process driven by an 
underlying Markov chain with transition probability matrix along the line segment
connecting $\bosy \Xi _1$ to $\bosy \Xi _2$. More specifically, it plots a
graph of 
$${\mathbf{E}}[A_1] = \sum_{i=1}^{3}p_{1}(i)\text{ }{\mathbf{E}}[A_1 \mid i],$$
where $p_{1}(i)$ is the steady state probability of state $i$ of a Markov chain
with transition probability matrix $\alpha \bosy \Xi _1 +  (1-\alpha) \bosy \Xi _2$,
as a function of the parameter $\alpha$. It can be seen that for an appropriate 
choice of ${\mathbf{E}}[B_1]$ (e.g., the one depicted by the horizontal dashed line)
$\bosy \Xi _1$ and $\bosy \Xi _2$ belong to ${\mathcal{I}}_{1}$ but there are points
on the line segment connecting $\bosy \Xi _1$ and $\bosy \Xi _2$ that don't.
Hence, in this case ${\mathcal{I}}_{1}$ is not convex.

\begin{figure}[ht]
\begin{center}
\mbox{\subfigure[\mbox{}]{
\psfrag{a}{\newline {\tiny $\alpha$}}
\psfrag{E[A1]}{\tiny ${\mathbf{E}}[A_1]$}
\includegraphics[width=0.45\textwidth,totalheight=0.35\textwidth]{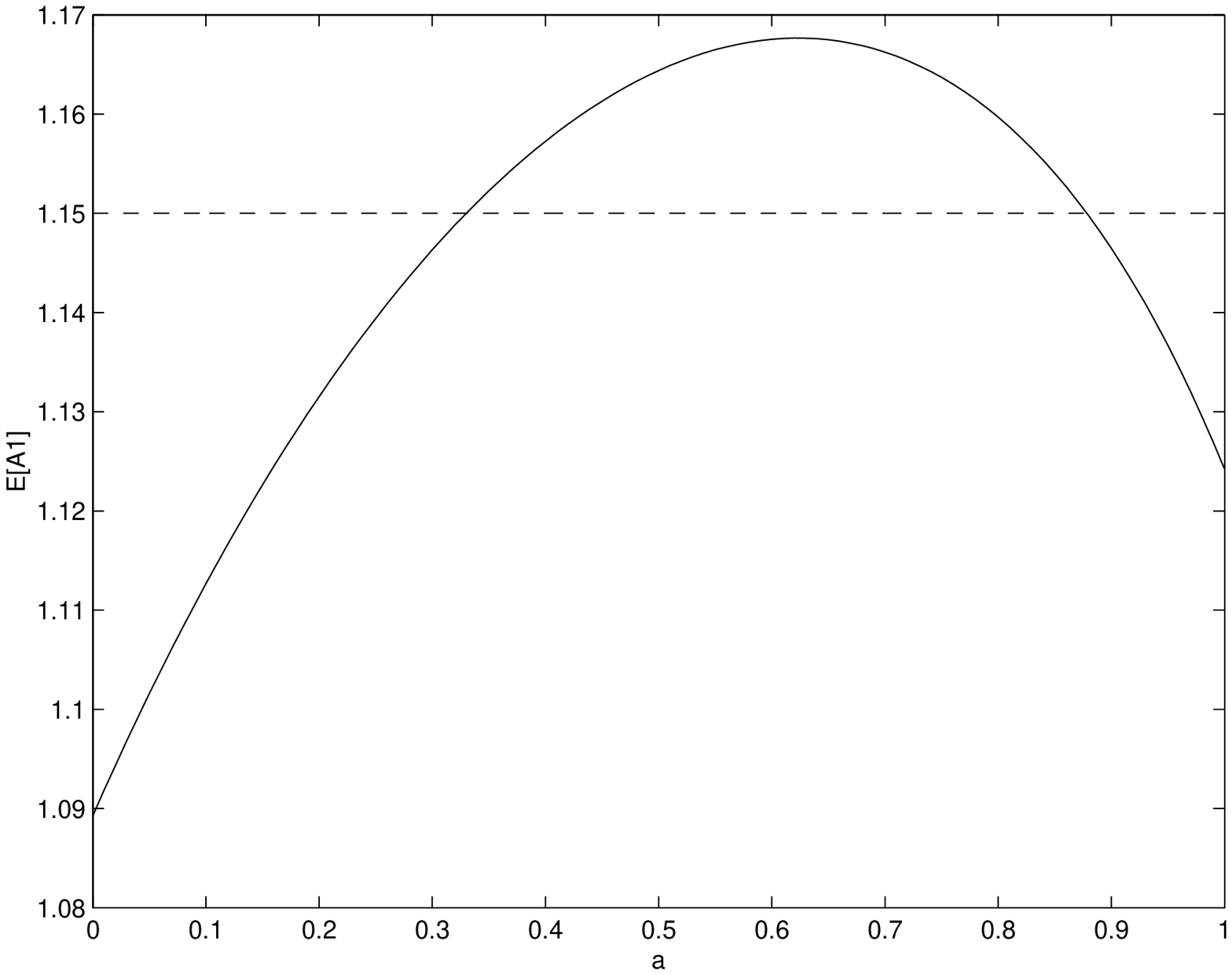}} 
\subfigure[\mbox{}]{
\psfrag{a}{\newline {\tiny $\alpha$}}
\psfrag{E[A1]}{\tiny ${\mathbf{E}}[A_1]$}
\includegraphics[width=0.45\textwidth,totalheight=0.35\textwidth]{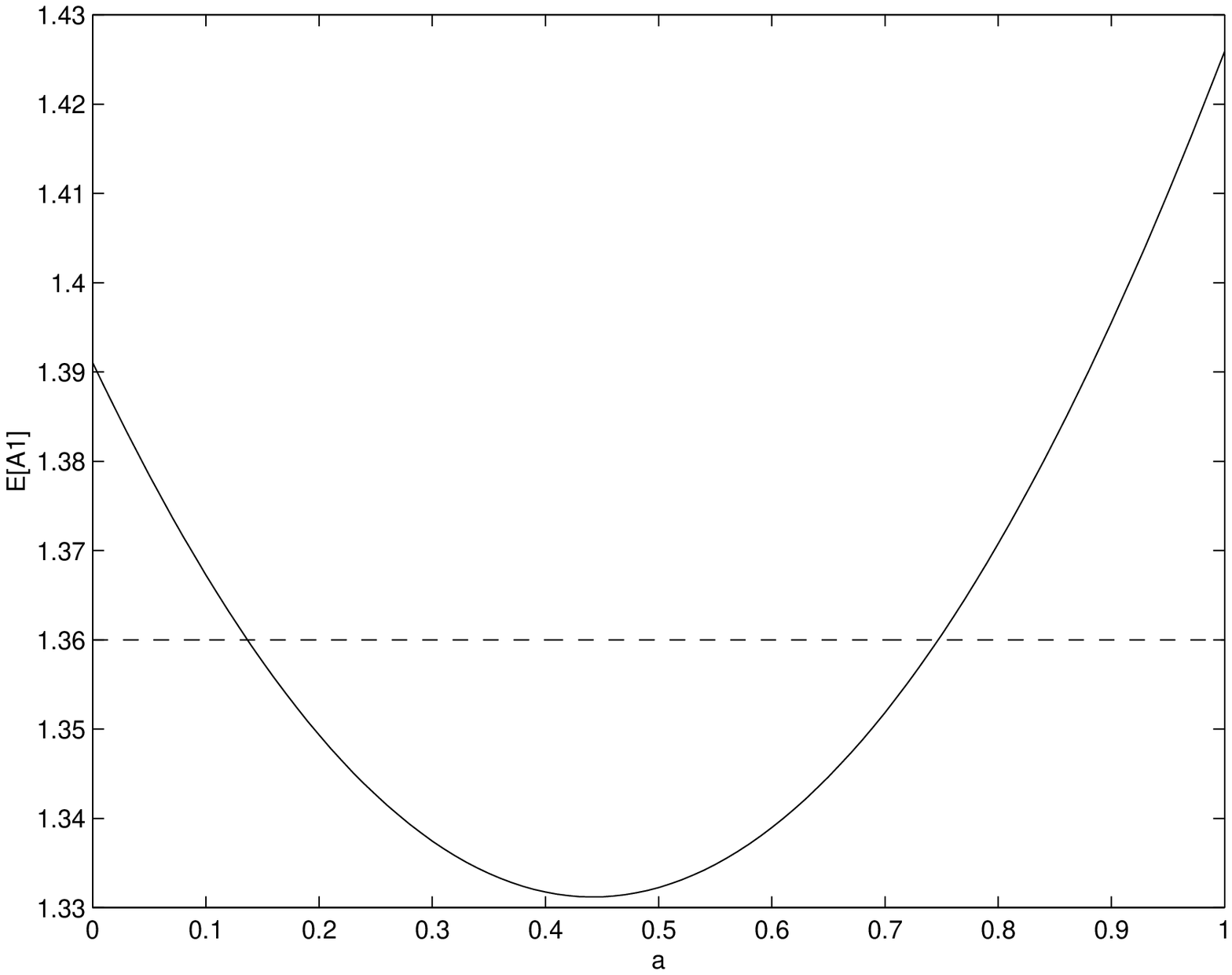}}}
\end{center}
\caption{Plots of ${\mathbf{E}}[A_1]$ along the line segments connecting two points in 
${\mathcal{I}}$ for examples where (a) ${\mathcal{I}}_{1}$ is not convex and (b) 
${\mathcal{I}}_{2}$ is not convex.}
\label{fig:noncon}
\end{figure}

Similarly, for a case where ${\mathcal{I}}_{2}$ is not convex, consider the 
transition probability matrices \\

\qquad $ {\bosy \Xi _1} =
\begin{bmatrix}
0.2 & 0.6 & 0.2 \\ 
0.2 & 0.2 & 0.6 \\ 
0.3 & 0.4 & 0.3 
\end{bmatrix} $
\qquad and \qquad
$ {\bosy \Xi _2} =
\begin{bmatrix}
0.2 & 0.1 & 0.7 \\ 
0.7 & 0.1 & 0.2 \\ 
0.4 & 0.2 & 0.4 
\end{bmatrix}$, \\
\\
and keep the same conditional mean arrival rates as in the previous example.
A plot of the mean arrival rate of the MMP along the line segment connecting 
$\bosy \Xi _1$ and $\bosy \Xi _2$ is shown in Figure~\ref{fig:noncon}(b).
Clearly, for an appropriate choice of ${\mathbf{E}}[B_1]$ (e.g., the one 
depicted by the horizontal dashed line)
$\bosy \Xi _1$ and $\bosy \Xi _2$ belong to ${\mathcal{I}}_{2}$ but there are points
on the line segment connecting $\bosy \Xi _1$ and $\bosy \Xi _2$ that don't.
Hence, in this case ${\mathcal{I}}_{2}$ is not convex.
  
In the next section we will see that $\bar{\theta}(%
{\mathbf{p}})$, the objective function in problem (\ref{NLP1-m}), is not
differentiable at ${\mathcal{I}}_{3}$. If ${\mathcal{I}}_{1}\,$was
always convex, we could take advantage of Lemma (\ref{sameSet}) and the
continuity of $\bar{\theta}({\mathbf{p}})$ to constraint the optimization into 
${\mathcal{I}}_{1}$. Unfortunately this is not the case. 

\section{Algorithms for solving the optimization problem} \label{sec:algo}

To solve the problem in (\ref{NLP1-m}) we have developed a heuristic algorithm
which performs very well in practice, typically giving first decimal digit
approximations to the optimal value by the third iteration and third decimal
digit by the fourth.
 
To describe the heuristic, note that the constraint set ${\mathcal{I}}$ is the 
Cartesian product of simplices, and that $q_f(j\mid i)>0$ implies $p(i,j)>0$ 
for all $i,j=1,\ldots,M$. This feasible set implies the following optimality
conditions (\cite[pg. 178-179]{bert-nlp})
\be{OptConds}
\frac{\partial \bar{\theta}({\mathbf{p}})}{\partial p(i,j)}=\frac{\partial 
\bar{\theta}({\mathbf{p}})}{\partial p(i,j')}\qquad \forall i,j,j'=1,\ldots,M,\  
\text{satisfying}\  q_f(j\mid i),q_f(j'\mid i)>0.  
\end{equation}
The heuristic algorithm iterates as follows:

{\bf Algorithm A}:
\begin{enumerate}
\item Initialize with ${\mb{p}}^{(0)}={\mb{q}}_f$ and m=0.
\item Form an approximation of $\bar{\theta}({\mb{p}})$ 
\be{heu-app} \bar{\theta}_{\text{app}}^{(m)}({\mb{p}}) \stackrel{\triangle}{=}
\ell s\theta^{*}({\mb{p}}^{(m)}) + \ell s \nabla\theta^{*}({\mb{p}}^{(m)})'
({\mb{p}}-{\mb{p}}^{(m)}) + \hat{I}_3({\mathbf{p}}). \end{equation}
Minimize the expression in Eq.~(\ref{heu-app}) subject to the constraints of
problem (\ref{NLP1-m}) to obtain an optimal solution
${\mathbf{p}}_{{\mathbf{p}}^{(m)}}^{*}$.
\item Set ${\mb{p}}^{(m+1)}:={\mathbf{p}}_{{\mathbf{p}}^{(m)}}^{*}$ and then make m=m+1. 
\item If $\bar{\theta}({\mathbf{p}}^{(m-1)})-\bar{\theta}({\mathbf{p}}^{(m)})<\epsilon$,
where $\epsilon$ is the desired accuracy, terminate with optimal solution ${\mb{p}}^{(m)}$. 
Otherwise, return to Step 2. 
\end{enumerate} 
The intuition behind this algorithm is that as we get closer to the real
minimum, ${\mb{p}}^*$, of problem (\ref{NLP1-m}) the approximations
$\bar{\theta}_{\text{app}}^{(m)}({\mb{p}})$ of $\bar{\theta}({\mathbf{p}})$
improve and yield ${\mb{p}}^{(m+1)}$ that are closer to ${\mb{p}}^*$.
Because at each step we solve the approximate problem exactly, the algorithm
typically needs much fewer iterations than a standard gradient-based
algorithm. 

The minimization of the expression in Eq.~(\ref{heu-app}) can be performed
by solving the system of optimality conditions: 
\begin{equation}
\frac{\partial \bar{\theta}_{app}^{(m)}({\mathbf{p}})}{\partial p(i,j)}=\frac{%
\partial \bar{\theta}_{app}^{(m)}({\mathbf{p}})}{\partial p(i,j')}\qquad
\forall i,j,j'=1,\ldots ,M\text{ satisfying }q_{f}(j\text{ }|\text{ }%
i),q_{f}(j'\text{ }|\text{ }i)>0  \label{OptCond2}
\end{equation}

which, by introducing the variables $w_i$, can be written equivalently as:
\begin{equation}
\frac{\partial \bar{\theta}_{app}^{(m)}({\mathbf{p}})}{\partial p(i,j)}=w
_{i}\qquad \forall i,j=1,\ldots ,M\text{ satisfying }q_{f}(j\text{ }|\text{ }%
i)>0  \label{OptCond31}
\end{equation}
subject to the conditions:
\begin{equation}
\sum_{j=1}^{M}p(i,j)=1\qquad \;\forall i=1,\ldots ,M  \label{OptCond32}
\end{equation}
and
\begin{equation}
p(i,j)>0 \qquad \;\forall i,j=1,\ldots ,M \text{ satisfying }q_{f}(j\text{ }|\text{ }i)>0 
\label{OptCond33}
\end{equation}

At the same time, $\forall i,j=1,\ldots ,M \text{ satisfying }q_{f}(j\text{ }|\text{ }%
i)=0$, $p(i,j)$ must be equal to zero as well.
Let us define the sets:
$$ {\mathcal{J}} (i) \triangleq \{ j=1,\ldots ,M \mid q_{f}(j\text{ }|\text{ }i)>0 \}. $$ 
The partial derivatives in Eq. (\ref{OptCond31}) are:
$$
\frac{\partial \bar{\theta}_{app}^{(m)}({\mathbf{p}})}{\partial p(i,j)}=\ell s%
\frac{\partial \theta ^{*}({\mathbf{p}}^{(m)})}{\partial p^{(m)}(i,j)}+\frac{%
\partial \widehat{I}_{3}({\mathbf{p}})}{\partial p(i,j)}
$$
where
\begin{equation}
\frac{\partial \widehat{I}_{3}({\mathbf{p}})}{\partial p(i,j)}=-\frac{%
q_{1}(i)q_{f}(j|i)}{p(i,j)} \label{I3parder}
\end{equation}

Therefore Eq. (\ref{OptCond31}) becomes:
\begin{equation}
\ell s\frac{\partial \theta ^{*}({\mathbf{p}}^{(m)})}{\partial p^{(m)}(i,j)}-\frac{%
q_{1}(i)q_{f}(j|i)}{p(i,j)}=w_{i}\qquad \forall i=1,\cdots ,M \text{ and } j \in {\mathcal{J}}(i)  
\end{equation}
which is equivalent to 
\be{solvedForpij}
p(i,j)=\frac{q_{1}(i)q_{f}(j|i)}{\ell s\frac{\partial \theta ^{*}({\mathbf{p}}^{(m)})}{\partial
p^{(m)}(i,j)}-w_{i}}\qquad \forall i=1,\cdots ,M \text{ and } j \in {\mathcal{J}}(i) 
\end{equation}

In order for the $p(i,j)$ to be positive we need:
$$ w_i < \ell s\frac{\partial \theta ^{*}({\mathbf{p}}^{(m)})}{\partial p^{(m)}(i,j)}\qquad
\forall i=1,\cdots ,M \text{ and } j \in {\mathcal{J}}(i) 
$$
or equivalently 
$$ w_i < \min_{j \in {\mathcal{J}}(i)} \ell s\frac{\partial \theta ^{*}({\mathbf{p}}^{(m)})}
{\partial p^{(m)}(i,j)} \triangleq w_{max} \qquad \forall i 
$$
We can then use Eq. (\ref{OptCond32}) to get:
$$
\sum_{j \in {\mathcal{J}}(i)}\frac{q_{1}(i)q_{f}(j|i)}{\ell s\frac{\partial \theta ^{*}({\mathbf{%
p}}^{(m)})}{\partial p^{(m)}(i,j)}-w_{i}}=1,\qquad \forall i
$$
The latest equation is a scalar equation in $w_{i}$ which can be solved
numerically in $(-\infty, w_{max})$. A unique solution is guaranteed since 
the function $\bar{\theta}_{app}^{(m)}({\mathbf{p}})$ is strictly convex and therefore 
has a unique minimum. We need to solve $M$ such equations one for each $w_{i}$.
Then by substituting the $w_{i}$'s back in Eq. (\ref{solvedForpij}) we
get all $p(i,j)$ that minimize the expression in Eq.~(\ref{heu-app}) to
form the optimal solution ${\mathbf{p}}_{{\mathbf{p}}^{(m)}}^{*}$. 

Although Algorithm A performs well in practice, it does not 
guarantee convergence. This is the case because the approximations
$\bar{\theta}_{\text{app}}^{(m)}({\mb{p}})$ are good in a small region around
the expansion point, but may be well off away from that point. This might lead
to unpleasant situations where $\bar{\theta}({\mathbf{p}}^{(m)})>
\bar{\theta}({\mathbf{p}}^{(m-1)})$. For the same reason, a step of the algorithm 
might produce a ${\mathbf{p}}^{(m)}$ that belongs to ${\mathcal{I}}_{2}$, 
independently of whether it is
improving the value of the objective function. In this case the next step
will yield ${\mathbf{p}}^{(m+1)}={\mathbf{q}}_{f}$, since $\bar{\theta}%
_{app}^{(m+1)}({\mathbf{p}})=\hat{I}_{3}({\mathbf{p}})$. Clearly, this will
bring the algorithm into an infinite loop. Lastly, we might be really unlucky
and land on a point ${\mathbf{p}}^{(m)} \in {\mathcal{I}}_{3}$. As we will see later on,
$\nabla\theta^{*}({\mb{p}}^{(m)})$  
and consequently $\bar{\theta}_{\text{app}}^{(m)}({\mb{p}})$ are not 
well defined at any ${\mathbf{p}}^{(m)} \in {\mathcal{I}}_{3}$ and hence the algorithm won't be able to continue.

Before we discuss an algorithm that does guarantee convergence, we need to
calculate the gradient of $\bar{\theta}({\mb{p}})$. The partial derivatives of 
$\hat{I}_3({\mathbf{p}})$ are given in Eq. (\ref{I3parder}) and are continuous
everywhere in ${\mathcal{I}}$. The calculation of the gradient of 
$\theta^*({\mb{p}})$ in ${\mathcal{I}}$ is a bit more involved:

Obviously, for ${\mathbf{p}} \in {\mathcal{I}}_{2}$, 
$\nabla\theta^{*}({\mathbf{p}}) = 0$.
In the following, we will calculate the gradient 
$\nabla\theta^{*}({\mathbf{p}})$, 
for  ${\mathbf{p}} \in {\mathcal{I}}_{1}$, i.e.,
in the region where $\theta^{*}({\mathbf{p}})>0$:

Let us denote by $\lambda_{i}$
the $i$-th eigenvalue of  $\bosy{\Pi}^A_{\theta,{\mb{p}}}$, and by ${\mathbf{u}}^{(i)}$ and
${\mathbf{v}}^{(i)}$ the (normalized)
left and right eigenvectors corresponding to this eigenvalue respectively (thus the spectral radius $\rho(\bosy{\Pi}^A_{\theta,{\mb{p}}}) = \lambda_1$).  Note that all the above eigenvalues and eigenvectors are functions of $\mb{p}$, $\theta$ and the marginal probability densities of arrival rates per state in the arrival process A. We will often omit these dependences for brevity. Assuming that the entries of  $\bosy{\Pi}^A_{\theta,{\mb{p}}}$ are functions of some parameter $\delta$ and since $\lambda_1$
has always multiplicity $1$ (Perron-Frobenius theorem), we have (see \cite{mur88} and \cite[Sec. 2.5-2.8]{wilk65}) 

\begin{equation}  \label{eig1}
\frac{\partial \lambda_1}{\partial \delta}=
\frac{{\mathbf{u}}^{(1)^T} (\frac{\partial}{\partial \delta} \bosy{\Pi}^A_{\theta,{\mb{p}}})  {\mathbf{v}}^{(1)}}{{\mathbf{u}}^{(1)^T}{\mathbf{v}}^{(1)}}. 
\end{equation}

Using the definition of $\bosy{\Pi}^A_{\theta,{\mb{p}}}$, i.e., $\bosy{\Pi}^A_{\theta,{\mb{p}}}=\{ \pi_{\theta, {\mb{p}}}^A(i,j)\}_{i,j=1,\ldots,M}^M \stackrel{\triangle}{=}
\{p(i,j) \eta_j(\theta)\}_{i,j=1,\ldots,M}^M$,
the above equation for $\delta = p(i,j)$ becomes:
\begin{equation}  \label{eig1b}
\frac{\partial \rho}{\partial p(i,j)}=
\frac{u_{i}^{(1)}v_{j}^{(1)}\eta_j(\theta)}{{\mathbf{u}}^{(1)^T}{\mathbf{v}}^{(1)}}. 
\end{equation}
and for $\delta = \theta$:
\begin{equation}  \label{eig1c}
\frac{\partial \rho}{\partial \theta}=
\sum_{i,j} \biggl[ \frac{u_{i}^{(1)}
v_{j}^{(1)}}{{\mathbf{u}}^{(1)^T} {\mathbf{v}}^{(1)}} p(i,j) \frac{\partial \eta_j(\theta)}{\partial \theta} \biggr]
\end{equation}

Now recall that
 $\Lambda_{A}(\theta,{\mathbf{p}})=\log \rho
(\bosy{\Pi}^A_{\theta,{\mb{p}}})$. Therefore:
\begin{equation} \label{eig2}
\frac{\partial \Lambda _{A}(\theta,{\mathbf{p}})}{\partial p(i,j)}=
\frac{\partial \log \rho(\bosy{\Pi}^A_{\theta,{\mb{p}}})}{\partial p(i,j)}=
\frac{1}{\rho(\bosy{\Pi}^A_{\theta,{\mb{p}}})} \frac{\partial \rho_1
(\bosy{\Pi}^A_{\theta,{\mb{p}}})}{\partial p(i,j)} 
= \frac{\eta_j(\theta)}{\rho(\bosy{\Pi}^A_{\theta,{\mb{p}}})}
\frac{u_{i}^{(1)} v_{j}^{(1)}}{{\mathbf{u}}^{(1)^T} {\mathbf{v}}^{(1)}}  
\end{equation}
Similarly, 
\bealn{eig2-1}
\frac{\partial \Lambda _{A}(\theta,{\mathbf{p}})}{\partial \theta} & = 
\frac{1}{\rho(\bosy{\Pi}^A_{\theta,{\mb{p}}})} \frac{\partial \rho
(\bosy{\Pi}^A_{\theta,{\mb{p}}})}{\partial \theta} \nonumber \\
& = \frac{1}{\rho(\bosy{\Pi}^A_{\theta,{\mb{p}}})} \sum_{i,j} \biggl[ \frac{u_{i}^{(1)} v_{j}^{(1)}}{{\mathbf{u}}^{(1)^T} {\mathbf{v}}^{(1)}}  p(i,j) \frac{\partial \eta_j(\theta)}{\partial \theta} \biggr]. 
\end{align}

Recall also that $\theta^*({\mb{p}})$ is the largest root of the equation
$\Lambda_A(\theta,{\mb{p}})+\Lambda_B(-\theta)=0$. Based on the discussion in
Section~\ref{sec:form},  $\theta^*({\mb{p}})$ can be alternatively written as
\begin{align}
\theta^*({\mb{p}}) = & \sup_{\{\theta\mid
\Lambda_A(\theta,{\mb{p}})+\Lambda_B(-\theta)<0\}} \theta \nonumber\\
= & \sup_{\{\theta\mid
\Lambda_A(\theta,{\mb{p}})+\Lambda_B(-\theta)\le0\}} \theta \label{eig3-2}\\ 
= & \inf_{a\ge0} \sup_\theta [ \theta - a \Lambda_A(\theta,{\mb{p}}) -a
\Lambda_B(-\theta) ],  \label{eig3-3}
\end{align}
where the second equality is due to the convexity of the limiting
$\log$-moment generating functions and the fact that
$\Lambda_A(\theta,{\mb{p}})+\Lambda_B(-\theta)$ is negative for sufficiently
small $\theta>0$. 
The third equality above is due to strong duality which holds since we
are dealing with a convex programming problem (see \cite[Chapter
5]{bert-nlp}).  
Using the envelope theorem \cite{var92}, we obtain
\be{eig4} \nabla \theta^*({\mb{p}})  = -a^* 
\nabla_{\mathbf{p}} \Lambda_A(\theta^*,{\mb{p}}), 
\end{equation}
where $a^*,\theta^*$ are optimal solutions of the optimization problems in
(\ref{eig3-3}), and the elements of $\nabla_{\mathbf{p}}
\Lambda_A(\theta^*,{\mb{p}})$ are given by (\ref{eig2}).  
Note that $a^*$ is a Lagrange multiplier for the original optimization problem
in (\ref{eig3-2})
and therefore satisfies the first order optimality condition
$$ 1 - 
a^* \frac{\partial (\Lambda_A(\theta^*,{\mb{p}})+\Lambda_B(-\theta^*))}{\partial
\theta^*} = 0, $$ 
which implies 
\be{eig5}
a^* = \biggl[
\frac{\partial (\Lambda_A(\theta^*,{\mb{p}})+\Lambda_B(-\theta^*))}{\partial
\theta^*}\biggr]^{-1}.
\end{equation}
Another way of deriving this result is by using the fact that
$\Lambda_A(\theta^*({\mb{p}}),{\mb{p}})+\Lambda_B(-\theta^*({\mb{p}}))$ 
is equal to zero for all ${\mb{p}}$ satisfying
$0<\theta^{*}({\mathbf{p}})<+\infty$. Thus, 
$$ \nabla_{\mathbf{p}}\Lambda_A(\theta^*({\mb{p}}),{\mb{p}})+
\nabla_{\mathbf{p}}\Lambda_B(-\theta^*({\mb{p}})) = 0, $$ 
and by using the chain rule we obtain
\be{eig6}
\nabla \theta^*({\mb{p}})  = -\frac{\nabla_{\mathbf{p}}
\Lambda_A(\theta^*,{\mb{p}})}{\frac{\partial}{\partial \theta^*}  
(\Lambda_A(\theta^*,{\mb{p}})+\Lambda_B(-\theta^*))},
\end{equation}
which is in agreement with (\ref{eig4})-(\ref{eig5}).

}

It is crucial to understand the behavior of $\nabla \bar{\theta}({\mathbf{p}})
$ and the directional derivatives of $\bar{\theta}({\mathbf{p}})$ along
feasible directions as ${\mathbf{p}}$ approaches the boundary ${\mathcal{I}}_{3}$
from within ${\mathcal{I}}_{1}$. We will denote this by ${\mathbf{p}}\stackrel{%
{\mathcal{I}}_{1}}{\rightarrow }{\mathbf{r}}$, where ${\mathbf{r}}$ is some
arbitrary point belonging to ${\mathcal{I}}_{3}$. Let us start by noting that
due to the continuity of $\theta ^{*}({\mathbf{p}})$, 
$$
\lim_{{\mathbf{p}}\stackrel{{\mathcal{I}}_{1}}{\rightarrow }{\mathbf{r}}}\theta
^{*}({\mathbf{p}})=\theta ^{*}({\mathbf{r}})=0
$$
On the same token, 
$$
\lim_{{\mathbf{p}}\stackrel{{\mathcal{I}}_{1}}{\rightarrow }{\mathbf{r}}}\eta
_{j}(\theta ^{*}({\mathbf{p}}))=\eta _{j}(0)=1
$$
which implies that:
$$
\lim_{{\mathbf{p}}\stackrel{{\mathcal{I}}_{1}}{\rightarrow }{\mathbf{r}}}{\mathbf{%
\Pi}}_{\theta ^{*},{\mathbf{p}}}^{A}=\lim_{{\mathbf{p}}\stackrel{{\mathcal{I}}_{1}%
}{\rightarrow }{\mathbf{r}}}\left\{ p(i,j)\eta _{j}(\theta ^{*}({\mathbf{p}}%
))\right\} =\left\{ p(i,j)\right\} =\bosy{\Xi }
$$
It is known that the spectral radius of $\bosy{\Xi }$, $\rho (\bosy{\Xi })=1$ and that its non-normalized right eigenvector has all its elements
equal to each other (the normalized one is $(1,1,$ $\ldots $ $,1)$) while its normalized left eigenvector comprises of the steady state probabilities for the Markov chain represented by $\bosy{\Xi }$. Thus:
$$
\lim_{{\mathbf{p}}\stackrel{{\mathcal{I}}_{1}}{\rightarrow }{\mathbf{r}}} \frac{u_{i}^{(1)} v_{j}^{(1)}}{{\mathbf{u}}^{(1)^T} {\mathbf{v}}^{(1)}} =\frac{u_{i}^{(1)}v_{j}^{(1)}}
{\sum\limits_{i}u_{i}^{(1)}v_{i}^{(1)}}=\frac{u_{i}^{(1)}v_{j}^{(1)}}{v_{j}^{(1)}\sum\limits_{i}u_{i}^{(1)}}=%
\frac{u_{i}^{(1)}}{\sum\limits_{i}u_{i}^{(1)}}=p_{1}(i)
$$
Furthermore, from the basic property of moment generating functions,
$$
\lim_{{\mathbf{p}}\stackrel{{\mathcal{I}}_{1}}{\rightarrow }{\mathbf{r}}}\left. 
\frac{\partial \eta _{j}(\theta )}{\partial \theta }\right| _{\theta
=\theta ^{*}({\mathbf{p}})}=\left. \frac{\partial \eta _{j}(\theta )}{%
\partial \theta }\right| _{\theta =0}={\mathbf{E}}[A_1 \mid j]
$$

Putting all these together we have that for all ${\mathbf{r}}\in {\mathcal{I}}_{3}$:

\begin{eqnarray*}
\lim_{{\mathbf{p}}\stackrel{{\mathcal{I}}_{1}}{\rightarrow }{\mathbf{r}}}\frac{%
\partial \Lambda _{A}(\theta ^{*},{\mathbf{p}})}{\partial \theta ^{*}}
&=&\lim_{{\mathbf{p}}\stackrel{{\mathcal{I}}_{1}}{\rightarrow }{\mathbf{r}}}\frac{1%
}{\rho ({\mathbf{\Pi}}_{\theta ^{*},{\mathbf{p}}}^{A})}\sum\limits_{i,j}\frac{%
u_{i}^{(1)}v_{j}^{(1)}}{{\mathbf{u}}^{(1)^T}{\mathbf{v}}^{(1)}}p(i,j)\frac{\partial \eta
_{j}(\theta ^{*})}{\partial \theta ^{*}}=\sum\limits_{i,j}p_{1}(i)p(i,j)%
{\mathbf{E}}[A_1 \mid j] \\
&=&\sum\limits_{j}{\mathbf{E}}[A_1 \mid j]\sum\limits_{i}p_{1}(i)p(i,j)=\sum%
\limits_{j}p_{1}(j){\mathbf{E}}[A_1 \mid j]={\mathbf{E}}[B_1]\text{ \quad }
\end{eqnarray*}

Moreover, 
$$
\frac {\partial \Lambda_B(-\theta^*)} {\partial \theta^*} = -{\mathbf{E}}[B_1]  
$$
 
Therefore the denominator in Eq. (\ref{eig6}) goes to zero for ${\mathbf{p}}%
\stackrel{{\mathcal{I}}_{1}}{\rightarrow }{\mathbf{r}}$. At the same time the
components of the numerator become:
$$
\lim_{{\mathbf{p}}\stackrel{{\mathcal{I}}_{1}}{\rightarrow }{\mathbf{r}}}\frac{%
\partial \Lambda _{A}(\theta ^{*},{\mathbf{p}})}{\partial p(i,j)}=\lim_{%
{\mathbf{p}}\stackrel{{\mathcal{I}}_{1}}{\rightarrow }{\mathbf{r}}}\frac{\eta
_{j}(\theta ^{*})}{\rho ({\mathbf{\Pi}}_{\theta ^{*},{\mathbf{p}}}^{A})}\frac{%
u_{i}^{(1)}v_{j}^{(1)}}{{\mathbf{u}}^{(1)^T}{\mathbf{v}}^{(1)}}=p_{1}(i)\neq 0
$$

We conclude that all partial derivatives tend to (plus or minus)\ infinity
as ${\mathbf{p}}\stackrel{{\mathcal{I}}_{1}}{\rightarrow }{\mathbf{r}}$. However,
this is not necessarily the case for directional derivatives along feasible
directions. Let us denote by ${\mathcal{D}}({\mathbf{p}})$ the set of all
feasible directions ${\mathbf{d}}$ at point ${\mathbf{p}}$. Note that all
feasible directions are of the form:

$${\mathbf{d}}=(d_{11},d_{12},\ldots
,d_{1M},\ldots ,d_{M1},d_{M2},\ldots d_{MM}) \text{ \qquad with  } \sum\limits_{j}d_{ij}=0, \quad \forall i.$$
In order to calculate the directional derivative we consider a normalized feasible direction vector which satisfies the additional constraint:
$$\sqrt{\sum\limits_{i,j}{d_{ij}^2}}=1, $$   
so that moving along the feasible direction is expressed as 
${\mathbf{p} + \beta \mathbf{d}}$ 

We denote by $\bosy{D}$ the $M \times M$ matrix corresponding to ${\mathbf{d}}$ and by $\bosy{H}$ the diagonal matrix with elements $\eta_i(\theta)$. We also denote by $\bosy{H}_{\theta}\stackrel{\triangle}{=}\frac{\partial \bosy{H}}{\partial\theta}$,  by $\bosy{H}_{\theta\theta}\stackrel{\triangle}{=}\frac{\partial^{2}\bosy{H}}{\partial\theta^{2}}$, by $\bosy{\Pi}_{\theta}\stackrel{\triangle}{=}\frac{\partial \bosy{\Pi}}{\partial\theta}$ and by $\bosy{\Pi}_{\beta}\stackrel{\triangle}{=}\frac{\partial \bosy{\Pi}}{\partial\beta}$. Note that:
$$ \bosy{\Pi}=\bosy{P}\cdot \bosy{H}\Rightarrow\left\{ \begin{array}{c}
\bosy{\Pi}_{\theta}=\bosy{P}\cdot \bosy{H}_{\theta}\\
\bosy{\Pi}_{\beta}=\bosy{D}\cdot \bosy{H}
\end{array}\right.
$$
Based on the above notation we can also write:
\begin{equation}  \label{dir_der_lambda_theta}
\frac{\partial {\Lambda}_{A}} {\partial \theta} = \frac{1}{\rho}
\frac{{\mathbf{u}}^{(1)^T} \bosy{P} \bosy{H}_{\theta}  {\mathbf{v}}^{(1)}}{{\mathbf{u}}^{(1)^T}{\mathbf{v}}^{(1)}}. 
\end{equation}

Now, the directional derivative of $\rho(\bosy{\Pi}^A_{\theta,{\mb{p}}})$ is derived from (\ref{eig1}) as:
\begin{equation}  \label{dir_der_rho}
\frac{\partial \lambda_1}{\partial \beta}=
\frac{{\mathbf{u}}^{(1)^T} (\frac{\partial}{\partial \beta} \bosy{\Pi}^A_{\theta,{\mb{p}}})  {\mathbf{v}}^{(1)}}{{\mathbf{u}}^{(1)^T}{\mathbf{v}}^{(1)}} = 
\frac{{\mathbf{u}}^{(1)^T} \bosy{D} \bosy{H}  {\mathbf{v}}^{(1)}}{{\mathbf{u}}^{(1)^T}{\mathbf{v}}^{(1)}}. 
\end{equation}

Therefore:
\begin{equation}  \label{dir_der_lambda_beta}
\frac{\partial {\Lambda}_{A}} {\partial \beta} = \frac{1}{\rho}
\frac{{\mathbf{u}}^{(1)^T} \bosy{D} \bosy{H}  {\mathbf{v}}^{(1)}}{{\mathbf{u}}^{(1)^T}{\mathbf{v}}^{(1)}}. 
\end{equation}

By the same arguments used above to obtain (\ref{eig6}) we obtain:
\be{dir_der_theta_beta}
\frac {\partial \theta^*} {\partial \beta}   = -\frac{ \frac{\partial {\Lambda}_{A}(\theta^*,{\mb{p}})} {{\partial \beta}} }{\frac{\partial}{\partial \theta^*}  
(\Lambda_A(\theta^*,{\mb{p}})+\Lambda_B(-\theta^*))} = 
-\frac{ \frac{1}{\rho}
\frac{{\mathbf{u}}^{(1)^T} \bosy{D} \bosy{H}  {\mathbf{v}}^{(1)}}{{\mathbf{u}}^{(1)^T}{\mathbf{v}}^{(1)}} }
{\frac{\partial}{\partial \theta^*}  
(\Lambda_A(\theta^*,{\mb{p}})+\Lambda_B(-\theta^*))},
\end{equation}

The denominator of this expression goes to zero for ${\mathbf{p}}\stackrel{%
{\mathcal{I}}_{1}}{\rightarrow }\mathbf{r}$, as discussed above. But the
numerator goes to zero as well, since:
$$
\lim_{{\mathbf{p}}\stackrel{{\mathcal{I}}_{1}}{\rightarrow }\mathbf{r}}
\frac{1}{\rho} \frac{{\mathbf{u}}^{(1)^T} \bosy{D} \bosy{H}  {\mathbf{v}}^{(1)}}{{\mathbf{u}}^{(1)^T}{\mathbf{v}}^{(1)}}  =
\lim_{{\mathbf{p}}\stackrel{{\mathcal{I}}_{1}}{\rightarrow }\mathbf{r}} \frac{1}{\rho}
\sum\limits_{i,j}d_{ij} \eta_{j}(\theta^{*})
\frac{u_{i}^{(1)}v_{j}^{(1)}}{{\mathbf{u}}^{(1)^T}{\mathbf{v}}^{(1)}}=\sum\limits_{i,j}
d_{ij}p_{1}(i)=\sum\limits_{i}p_{1}(i)\sum
\limits_{j}d_{ij}=0
$$

It can be proven using De L' H$\hat{o}$pital's theorem that the limit of all directional derivatives
towards directions pointing into ${\mathcal{I}}_{1}$ as ${\mathbf{p}}\stackrel{{\mathcal{I}}_{1}}{\rightarrow }\mathbf{r}$ is zero. However, the
proof is quite lengthy and is omitted due to space limitations.

Thus, the necessary and sufficient (due to convexity) condition for a point 
${\mathbf{p}}^{*}\in {\mathcal{I}}$ to be the global minimum of $\bar{\theta}(%
{\mathbf{p}})\,$can be written as:
$$
\lim_{a\downarrow 0}\frac{\bar{\theta}({\mathbf{p}}^{*}+a{\mathbf{d}})-
\bar{\theta}({\mathbf{p}}^{*})}{a}\geq 0, \qquad \forall {\mathbf{d}}\in 
{\mathcal{D}}({\mathbf{p}})
$$

\begin{lemma} \label{gradRel}
The direction sequence $\{{\mathbf{d}}^{(m)}\}=\{{\mathbf{p}}_{%
{\mathbf{p}}^{(m)}}^{*}-{\mathbf{p}}^{(m)}\}$ is gradient related to $\{\mathbf{p%
}^{(m)}\}$ in ${\mathcal{I}}_{1}\cup {\mathcal{I}}_{2}$, that is for any
subsequence $\{{\mathbf{p}}^{(m)}\}_{m\in \mathcal{M}}$ such that ${\mathbf{p}}%
^{(m)}\notin {\mathcal{I}}_{3}$, that converges to a non-stationary point $%
{\mathbf{p}}^{(0)}\in {\mathcal{I}}$ of $\bar{\theta}({\mathbf{p}})$, the
corresponding subsequence $\{{\mathbf{d}}^{(m)}\}_{m\in \mathcal{M}}\,$ is
bounded and satisfies 
$$
\limsup_{m\rightarrow \infty ,\text{ }m\in \mathcal{M}} \nabla 
\bar{\theta}({\mathbf{p}}^{(m)})^{\prime }{\mathbf{d}}^{(m)}<0
$$
\end{lemma}

\pf By definition, ${\mathbf{p}}_{{\mathbf{p}}^{(m)}}^{*} \in {\mathcal{I}}$ which 
is bounded. Therefore, $\{{\mathbf{d}}^{(m)}\}$ is bounded, too.

Note that for all ${\mathbf{p}}^{(m)}\mathbf{,}$ $\nabla \bar{\theta}(\mathbf{p%
}^{(m)})=\nabla \bar{\theta}_{{\mathbf{p}}^{(m)}}^{app}({\mathbf{p}}^{(m)})$.

By definition:
\begin{equation}
\nabla \bar{\theta}_{{\mathbf{p}}^{(m)}}^{app}({\mathbf{p}}^{(m)})^{\prime }(%
{\mathbf{p}}_{{\mathbf{p}}^{(m)}}^{*}-{\mathbf{p}}^{(m)})=\lim_{a\downarrow 0}%
\frac{\bar{\theta}_{{\mathbf{p}}^{(m)}}^{app}({\mathbf{p}}^{(m)}+a({\mathbf{p}}_{%
{\mathbf{p}}^{(m)}}^{*}-{\mathbf{p}}^{(m)}))-\bar{\theta}_{{\mathbf{p}}%
^{(m)}}^{app}({\mathbf{p}}^{(m)})}{a}  \label{DirDer}
\end{equation}

Now note that $\bar{\theta}_{{\mathbf{p}}^{(m)}}^{app}(\cdot )$ is a strictly
convex function whose global minimum is ${\mathbf{p}}_{{\mathbf{p}}^{(m)}}^{*}$.
Thus,
\bealn{uppBound}
\frac{\bar{\theta}_{{\mathbf{p}}^{(m)}}^{app}({\mathbf{p}}^{(m)}+a({\mathbf{p}}_{%
{\mathbf{p}}^{(m)}}^{*}-{\mathbf{p}}^{(m)}))-\bar{\theta}_{{\mathbf{p}}%
^{(m)}}^{app}({\mathbf{p}}^{(m)})}{a} = \\ \nonumber \frac{\bar{\theta}_{{\mathbf{p}}%
^{(m)}}^{app}((1-a){\mathbf{p}}^{(m)}+a{\mathbf{p}}_{{\mathbf{p}}^{(m)}}^{*})-\bar{%
\theta}_{{\mathbf{p}}^{(m)}}^{app}({\mathbf{p}}^{(m)})}{a}< \\ \nonumber
\frac{(1-a)\bar{\theta}_{{\mathbf{p}}^{(m)}}^{app}({\mathbf{p}}^{(m)})+a\bar{%
\theta}_{{\mathbf{p}}^{(m)}}^{app}({\mathbf{p}}_{{\mathbf{p}}^{(m)}}^{*})-\bar{%
\theta}_{{\mathbf{p}}^{(m)}}^{app}({\mathbf{p}}^{(m)})}{a}= \\ \nonumber 
\frac{-a\bar{\theta}_{{\mathbf{p}}^{(m)}}^{app}({\mathbf{p}}^{(m)})+a\bar{\theta}_{{\mathbf{p}}%
^{(m)}}^{app}({\mathbf{p}}_{{\mathbf{p}}^{(m)}}^{*})}{a}= \bar{\theta}_{{\mathbf{p}}%
^{(m)}}^{app}({\mathbf{p}}_{{\mathbf{p}}^{(m)}}^{*})-\bar{\theta}_{{\mathbf{p}}%
^{(m)}}^{app}({\mathbf{p}}^{(m)})
\end{align}

Equations (\ref{DirDer}) and (\ref{uppBound}) imply that:
$$
\nabla \bar{\theta}_{{\mathbf{p}}^{(m)}}^{app}({\mathbf{p}}^{(m)})^{\prime }(%
{\mathbf{p}}_{{\mathbf{p}}^{(m)}}^{*}-{\mathbf{p}}^{(m)})\leq \bar{\theta}_{%
{\mathbf{p}}^{(m)}}^{app}({\mathbf{p}}_{{\mathbf{p}}^{(m)}}^{*})-\bar{\theta}_{%
{\mathbf{p}}^{(m)}}^{app}({\mathbf{p}}^{(m)}) 
$$
Therefore:
$$
\limsup_{m\rightarrow \infty ,\text{ }m\in \mathcal{M}} \nabla 
\bar{\theta}_{{\mathbf{p}}^{(m)}}^{app}({\mathbf{p}}^{(m)})^{\prime }({\mathbf{p}}%
_{{\mathbf{p}}^{(m)}}^{*}-{\mathbf{p}}^{(m)})\leq \limsup_{m\rightarrow
\infty ,\text{ }m\in \mathcal{M}}{\text{ }}\bar{\theta}_{{\mathbf{p}}%
^{(m)}}^{app}({\mathbf{p}}_{{\mathbf{p}}^{(m)}}^{*})-\bar{\theta}_{{\mathbf{p}}%
^{(m)}}^{app}({\mathbf{p}}^{(m)})
$$
We now distinguish between two cases: ${\mathbf{p}}^{(0)}\in {\mathcal{I}}_{3%
\text{ }}$and ${\mathbf{p}}^{(0)}\notin {\mathcal{I}}_{3}$.

If ${\mathbf{p}}^{(0)}\notin {\mathcal{I}}_{3}$ then by continuity arguments we
can conclude that:
$$
\limsup_{m\rightarrow \infty ,\text{ }m\in \mathcal{M}}{\text{ }}
\bar{\theta}_{{\mathbf{p}}^{(m)}}^{app}({\mathbf{p}}_{{\mathbf{p}}^{(m)}}^{*})-%
\bar{\theta}_{{\mathbf{p}}^{(m)}}^{app}({\mathbf{p}}^{(m)})=\bar{\theta}_{%
{\mathbf{p}}^{(0)}}^{app}({\mathbf{p}}_{{\mathbf{p}}^{(0)}}^{*})-\bar{\theta}_{%
{\mathbf{p}}^{(0)}}^{app}({\mathbf{p}}^{(0)})
$$
But $\bar{\theta}_{{\mathbf{p}}^{(0)}}^{app}(\cdot )$ is also a strictly
convex function whose global minimum is ${\mathbf{p}}_{{\mathbf{p}}^{(0)}}^{*}$
and therefore unless ${\mathbf{p}}^{(0)}={\mathbf{p}}_{{\mathbf{p}}^{(0)}}^{*}$
(in which case ${\mathbf{p}}^{(0)}$ is the global minimum of both $\bar{\theta}%
_{{\mathbf{p}}^{(0)}}^{app}(\cdot )$ and $\bar{\theta}(\cdot )$), $\bar{\theta}%
_{{\mathbf{p}}^{(0)}}^{app}({\mathbf{p}}_{{\mathbf{p}}^{(0)}}^{*})-\bar{\theta}_{%
{\mathbf{p}}^{(0)}}^{app}({\mathbf{p}}^{(0)})<0$. 

If ${\mathbf{p}}^{(0)}\in {\mathcal{I}}_{3}$, we have three sub-cases:
\begin{enumerate}

\item  $\exists m_{1}\in \mathcal{M}$ s.t. $\forall m\in \mathcal{M}$ with $%
m>m_{1}$, ${\mathbf{p}}^{(m)}\in {\mathcal{I}}_{2}$.

In this case $\forall m\in \mathcal{M}$ with $m>m_{1}\,$we have ${\mathbf{p}}_{%
{\mathbf{p}}^{(m)}}^{*}=\mathbf{q}_{f}$ which yields $\bar{\theta}_{{\mathbf{p}}%
^{(m)}}^{app}({\mathbf{p}}_{{\mathbf{p}}^{(m)}}^{*})=0$ and $\bar{\theta}_{%
{\mathbf{p}}^{(m)}}^{app}({\mathbf{p}}^{(m)})=\widehat{I}_{3}({\mathbf{p}}^{(m)})$. 
Hence:
$$
\limsup_{m\rightarrow \infty ,\text{ }m\in \mathcal{M}}{\text{ }}
\bar{\theta}_{{\mathbf{p}}^{(m)}}^{app}({\mathbf{p}}_{{\mathbf{p}}^{(m)}}^{*})-%
\bar{\theta}_{{\mathbf{p}}^{(m)}}^{app}({\mathbf{p}}^{(m)})=-\widehat{I}_{3}(%
{\mathbf{p}}^{(0)})
$$
But this is always less than zero, since $\mathbf{q}_{f}$ $\in {\mathcal{I}}%
_{1}$ and ${\mathbf{p}}^{(0)}\in {\mathcal{I}}_{3}$.

\item  $\exists m_{2}\in \mathcal{M}$ s.t. $\forall m\in \mathcal{M}$ with $%
m>m_{2}$, ${\mathbf{p}}^{(m)}\in {\mathcal{I}}_{1}$.

Consider a function $\bar{\varphi}({\mathbf{p}})$ which is equal to $\bar{%
\theta}({\mathbf{p}})$, $\forall {\mathbf{p}}\in {\mathcal{I}}_{4}={\mathcal{I}}_{1}\cup {\mathcal{I}}%
_{3}$ and continuously differentiable at any $\mathbf{r\in }{\mathcal{I}}_{3}$%
. Define $\bar{\varphi}_{{\mathbf{p}}^{(m)}}^{app}({\mathbf{p}})$ in the
same way that $\bar{\theta}_{{\mathbf{p}}^{(m)}}^{app}({\mathbf{p}})$ is
defined. Then:
$$
\limsup_{m\rightarrow \infty ,\text{ }m\in \mathcal{M}}{\text{ }}
\bar{\theta}_{{\mathbf{p}}^{(m)}}^{app}({\mathbf{p}}_{{\mathbf{p}}^{(m)}}^{*})-%
\bar{\theta}_{{\mathbf{p}}^{(m)}}^{app}({\mathbf{p}}^{(m)})=\bar{\varphi}_{%
{\mathbf{p}}^{(0)}}^{app}({\mathbf{p}}_{{\mathbf{p}}^{(0)}}^{*})-\bar{\varphi}_{%
{\mathbf{p}}^{(0)}}^{app}({\mathbf{p}}^{(0)})
$$
But $\bar{\varphi}_{{\mathbf{p}}^{(m)}}^{app}({\mathbf{p}})$ is also strictly
convex (by definition) and therefore $\bar{\varphi}_{{\mathbf{p}}^{(0)}}^{app}(%
{\mathbf{p}}_{{\mathbf{p}}^{(0)}}^{*})-\bar{\varphi}_{{\mathbf{p}}^{(0)}}^{app}(%
{\mathbf{p}}^{(0)})<0$ unless ${\mathbf{p}}^{(0)}$ is the global minimum of $%
\bar{\varphi}_{{\mathbf{p}}^{(m)}}^{app}({\mathbf{p}})$. In that case ${\mathbf{p}}%
^{(0)}$ would also be the global minimum of $\bar{\varphi}({\mathbf{p}})$
which in turn would make it the global minimum of $\bar{\theta}({\mathbf{p}})$
in ${\mathcal{I}}_{4}$. But as seen in the proof of Lemma
\ref{sameSet} the global minimum of $\bar{\theta}({\mathbf{p}})$ is always
in ${\mathcal{I}}_{4}$ which means that ${\mathbf{p}}^{(0)}$
should be the global minimum of $\bar{\theta}({\mathbf{p}})$ in ${\mathcal{I}}$.
This clearly contradicts our assumption that ${\mathbf{p}}^{(0)}$ is not a
stationary point of $\bar{\theta}({\mathbf{p}})$. Hence, $\bar{\varphi}_{%
{\mathbf{p}}^{(0)}}^{app}({\mathbf{p}}_{{\mathbf{p}}^{(0)}}^{*})-\bar{\varphi}_{%
{\mathbf{p}}^{(0)}}^{app}({\mathbf{p}}^{(0)})$ is always less than $0$.

\item  ${\mathcal{M}}$ can be partitioned into two unbounded sets ${\mathcal{M}}%
_{1}$ and ${\mathcal{M}}_{2}$ with ${\mathbf{p}}^{(m)}\in {\mathcal{I}}_{1}$, $%
\forall m\in {\mathcal{M}}_{1}$ and ${\mathbf{p}}^{(m)}\in {\mathcal{I}}_{2}$, $%
\forall m\in {\mathcal{M}}_{2}$. Then:
\beal{}
\limsup_{m\rightarrow \infty ,\text{ }m\in \mathcal{M}}
\bar{\theta}_{{\mathbf{p}}^{(m)}}^{app}({\mathbf{p}}_{{\mathbf{p}}^{(m)}}^{*})-
\bar{\theta}_{{\mathbf{p}}^{(m)}}^{app}({\mathbf{p}}^{(m)})= \\ \nonumber
=\max [\limsup_{m\rightarrow \infty ,\text{ }m\in {\mathcal{M}}_{1}}
\bar{\theta}_{{\mathbf{p}}^{(m)}}^{app}({\mathbf{p}}_{{\mathbf{p}}^{(m)}}^{*})
-\bar{\theta}_{{\mathbf{p}}^{(m)}}^{app}({\mathbf{p}}^{(m)}), \\ \nonumber
\limsup_{m\rightarrow \infty ,\text{ }m\in {\mathcal{M}}_{2}} 
\bar{\theta}_{{\mathbf{p}}^{(m)}}^{app}({\mathbf{p}}_{{\mathbf{p}}^{(m)}}^{*})
-\bar{\theta}_{{\mathbf{p}}^{(m)}}^{app}({\mathbf{p}}^{(m)}) ] 
\end{align*}

But each one of these two limits superior belongs to the special cases
analyzed above. Therefore, each one is less than zero and the maximum is
always less than zero.
\end{enumerate}

This completes the proof.
\qed

Lemma~\ref{gradRel} can be used to prove that the following algorithm always
converges:

\textbf{Algorithm B}:

\begin{enumerate}
\item[1-3.]  Steps 1-3 are exactly the same as in Algorithm A.

\item[4.]  Set ${\mathbf{p}}^{(m+1)}={\mathbf{p}}^{(m)}+a^{(m)}({\mathbf{p}}_{%
{\mathbf{p}}^{(m)}}^{*}-{\mathbf{p}}^{(m)})$ where $a^{(m)}$ is the stepsize.
Use a slightly modified Armijo rule \cite{bert-nlp} to select the
stepsize. To implement this rule, fixed scalars $\beta \in (0,1)$ and $%
\sigma $ $\in (0,1)$ are chosen and we set $a^{(m)}=\beta ^{l^{\prime }}$,
where $l^{\prime }$ is the first nonnegative integer $l$ for which
$$
{\mathbf{p}}^{(m)}+\beta ^{l}({\mathbf{p}}_{{\mathbf{p}}^{(m)}}^{*}-{\mathbf{p}}%
^{(m)})\notin {\mathcal{I}}_{3}
$$
and
$$
\bar{\theta}({\mathbf{p}}^{(m)})-\bar{\theta}\left( {\mathbf{p}}^{(m)}+\beta
^{l}({\mathbf{p}}_{{\mathbf{p}}^{(m)}}^{*}-{\mathbf{p}}^{(m)})\right) \geq -\sigma
\beta ^{l}\nabla \bar{\theta}({\mathbf{p}}^{(m)})^{\prime }({\mathbf{p}}_{%
{\mathbf{p}}^{(m)}}^{*}-{\mathbf{p}}^{(m)})
$$
Such a stepsize is guaranteed to exist since the set of stepsizes that
satisfy the above inequality will always contain an interval of the form $%
[0,\delta ]\,$with $\delta >0$ and ${\mathbf{p}}^{(m)}\notin {\mathcal{I}}%
_{3}\Rightarrow \exists \varepsilon >0$ s.t. $\forall 0\leq a<\varepsilon
,\quad {\mathbf{p}}^{(m)}+a({\mathbf{p}}_{{\mathbf{p}}^{(m)}}^{*}-{\mathbf{p}}%
^{(m)})\notin {\mathcal{I}}_{3}$. Return to step 2.
\end{enumerate}

Note that for $l=0$, ${\mathbf{p}}^{(m+1)}={\mathbf{p}}_{{\mathbf{p}}^{(m)}}^{*}$
and therefore if the above conditions are satisfied at first trial,
Algorithm B behaves exactly as Algorithm A. Thus, Algorithm B guarantees
convergence without sacrificing much in performance.
\begin{thm} \label{thm:Convergence}
Let $\{{\mathbf{p}}^{(m)}\}$ be a sequence generated by Algorithm
B. Then $\{{\mathbf{p}}^{(m)}\}$ converges to ${\mathbf{p}}^{*}$, the global
minimum of $\bar{\theta}({\mathbf{p}})$. 
\end{thm}
\pf (based on the proof of Proposition 1.2.1 in \cite{bert-nlp}):
 
Assume that ${\mathbf{p}}_{0}$ is a limit point of $\{%
{\mathbf{p}}^{(m)}\}$ that is not the global minimum, i.e.,
\begin{equation}
\exists {\mathbf{d}}\in {\mathcal{D}}({\mathbf{p}})\text{ s.t.\qquad }%
\lim_{a\downarrow 0}\frac{\bar{\theta}({\mathbf{p}}_{0}+a{\mathbf{d}})-\bar{%
\theta}({\mathbf{p}}_{0})}{a}<0  \label{nonSmoothOptCond}
\end{equation}

Note that since $\{\bar{\theta}({\mathbf{p}}^{(m)})\}$ is monotonically
non-increasing, it either converges to a finite value or diverges to -$\infty 
$. Since $\bar{\theta}({\mathbf{p}})$ is continuous, $\bar{\theta}({\mathbf{p}}%
_{0})$ is a limit point of $\{\bar{\theta}({\mathbf{p}}^{(m)})\}$ and it
follows that the entire sequence $\{\bar{\theta}({\mathbf{p}}^{(m)})\}$
converges to $\bar{\theta}({\mathbf{p}}_{0})$. Hence, $\bar{\theta}({\mathbf{p}}%
^{(m)})-\bar{\theta}({\mathbf{p}}^{(m+1)})\rightarrow 0$. 

By the definition of the stepsize selection rule in Algorithm B, we have
that all ${\mathbf{p}}^{(m)}\notin {\mathcal{I}}_{3}$ (although ${\mathbf{p}}_{0}$
can belong to ${\mathcal{I}}_{3}$) and:
$$
\bar{\theta}({\mathbf{p}}^{(m)})-\bar{\theta}({\mathbf{p}}^{(m+1)})\geq -\sigma
a^{(m)}\nabla \bar{\theta}({\mathbf{p}}^{(m)})^{\prime }{\mathbf{d}}^{(m)}
$$
Hence, $a^{(m)}\nabla \bar{\theta}({\mathbf{p}}^{(m)})^{\prime }{\mathbf{d}}%
^{(m)}\rightarrow 0$. Let $\{{\mathbf{p}}^{(m)}\}_{\mathcal{M}}$ be a
subsequence converging to ${\mathbf{p}}_{0}$. From Lemma~\ref{gradRel}, we have
$$
\limsup_{m\rightarrow \infty ,\text{ }m\in \mathcal{M}} \nabla 
\bar{\theta}({\mathbf{p}}^{(m)})^{\prime }{\mathbf{d}}^{(m)}<0
$$
and therefore $\{a^{(m)}\}_{\mathcal{M}}\rightarrow 0$.

Now, by the definition of the stepsize selection rule and the fact that $%
\{a^{(m)}\}_{\mathcal{M}}$ goes to zero, there must be some $\bar{m}$ such that
the initial stepsize $s$ will be reduced for at least two times for all $%
m\in \mathcal{M},$ $m\geq \bar{m}$. But due to the shape and dimension of
${\mathcal{I}}_{3}$, its intersection with the line segment connecting 
${\mathbf{p}}^{(m)}$ and ${\mathbf{p}}_{{\mathbf{p}}^{(m)}}^{*}$ can be at most
a single point. Therefore, at most one of the stepsize reductions occurs
because ${\mathbf{p}}^{(m)}+\beta ^{l}({\mathbf{p}}_{{\mathbf{p}}^{(m)}}^{*}-%
{\mathbf{p}}^{(m)})\in {\mathcal{I}}_{3}$, and at least one of the following
two inequalities holds:
\begin{equation}
\bar{\theta}({\mathbf{p}}^{(m)})-\bar{\theta}({\mathbf{p}}^{(m)}+(a^{(m)}/\beta )%
{\mathbf{d}}^{(m)})<-\sigma (a^{(m)}/\beta )\nabla \bar{\theta}({\mathbf{p}}%
^{(m)})^{\prime }{\mathbf{d}}^{(m)}  \label{setM1}
\end{equation}
or
\begin{equation}
\bar{\theta}({\mathbf{p}}^{(m)})-\bar{\theta}({\mathbf{p}}^{(m)}+(a^{(m)}/\beta
^{2}){\mathbf{d}}^{(m)})<-\sigma (a^{(m)}/\beta ^{2})\nabla \bar{\theta}(%
{\mathbf{p}}^{(m)})^{\prime }{\mathbf{d}}^{(m)}  \label{setM2}
\end{equation}

Let us denote by ${\mathcal{M}}_{1}$ the set of $m$'s for which 
Eq.~(\ref{setM1}) holds and by ${\mathcal{M}}_{2}$, the set of $m$'s for which 
Eq.~(\ref{setM1}) does not hold and therefore Eq.~(\ref{setM2}) does
hold. Define:
$$
{\mathbf{x}}^{(m)}=\frac{{\mathbf{d}}^{(m)}}{\left\| {\mathbf{d}}^{(m)}\right\| }
$$
and
$$
\bar{a}^{(m)}=\left\{ 
\begin{array}{ll}
\frac{a^{(m)}\left\| {\mathbf{d}}^{(m)}\right\| }{\beta }\text{ \qquad if }m\in 
{\mathcal{M}}_{1} \\ \\
\frac{a^{(m)}\left\| {\mathbf{d}}^{(m)}\right\| }{\beta ^{2}}\text{ \qquad if 
}m\in {\mathcal{M}}_{2}
\end{array}
\right.
$$

Since $\{{\mathbf{d}}^{(m)}\}$ is gradient related, $\{\left\| {\mathbf{d}}%
^{(m)}\right\| \}_{\mathcal{M}}$ is bounded and therefore $\{\bar{a}%
^{(m)}\}_{\mathcal{M}}\rightarrow 0$. Since $\left\| {\mathbf{x}}%
^{(m)}\right\| =1$ for all $m\in \mathcal{M}$, there exists a subsequence $\{%
{\mathbf{x}}^{(m)}\}_{\mathcal{\bar{M}}}$ of $\{{\mathbf{x}}^{(m)}\}_{\mathcal{M}%
}$ such that $\{{\mathbf{x}}^{(m)}\}_{\mathcal{\bar{M}}}\rightarrow \mathbf{%
\bar{x}}$ where $\mathbf{\bar{x}\,}$is some vector with $\left\| \mathbf{%
\bar{x}}\right\| =1$ (see Prop. A.5(c) in Appendix A\ of \cite{bert-nlp}). 

Given the above definitions, Eq. (\ref{setM1}) and (\ref{setM2}) can be
combined into:
$$
\frac{\bar{\theta}({\mathbf{p}}^{(m)})-\bar{\theta}({\mathbf{p}}^{(m)}+\bar{a}%
^{(m)}{\mathbf{x}}^{(m)})}{\bar{a}^{(m)}}<-\sigma \nabla \bar{\theta}({\mathbf{p%
}}^{(m)})^{\prime }{\mathbf{x}}^{(m)},\qquad \forall m\in {\mathcal{\bar{M}}},
m\geq \bar{m}
$$
Since $\bar{\theta}({\mathbf{p}})$ is a convex and continuous function there
must be some $\tilde{a}^{(m)}\in [0,\bar{a}^{(m)}]$ with ${\mathbf{p}}^{(m)}+%
\tilde{a}^{(m)}{\mathbf{x}}^{(m)}\notin {\mathcal{I}}_{3}$ such that:
\beal
-\nabla \bar{\theta}({\mathbf{p}}^{(m)}+\tilde{a}^{(m)}{\mathbf{x}}%
^{(m)})^{\prime }{\mathbf{x}}^{(m)} & \leq \frac{\bar{\theta}({\mathbf{p}}^{(m)})-%
\bar{\theta}({\mathbf{p}}^{(m)}+\bar{a}^{(m)}{\mathbf{x}}^{(m)})}{\bar{a}^{(m)}} \\ \nonumber
& <-\sigma \nabla \bar{\theta}({\mathbf{p}}^{(m)})^{\prime }{\mathbf{x}}%
^{(m)},\qquad \forall m\in {\mathcal{\bar{M}}}, m\geq \bar{m} 
\end{align*}
Taking limits in the above equation we obtain
$$
-\lim_{a\downarrow 0}\frac{\bar{\theta}({\mathbf{p}}_{0}+a\mathbf{\bar{x}})-%
\bar{\theta}({\mathbf{p}}_{0})}{a}\leq -\sigma \lim_{a\downarrow 0}\frac{\bar{%
\theta}({\mathbf{p}}_{0}+a\mathbf{\bar{x}})-\bar{\theta}({\mathbf{p}}_{0})}{a}
$$
or 
$$
0\leq (1-\sigma )\lim_{a\downarrow 0}\frac{\bar{\theta}({\mathbf{p}}_{0}+a%
\mathbf{\bar{x}})-\bar{\theta}({\mathbf{p}}_{0})}{a}
$$
And since $\sigma <1$, it follows that 
$$
\lim_{a\downarrow 0}\frac{\bar{\theta}({\mathbf{p}}_{0}+a\mathbf{\bar{x}})-%
\bar{\theta}({\mathbf{p}}_{0})}{a}\geq 0
$$
But this contradicts Eq. (\ref{nonSmoothOptCond}) and therefore ${\mathbf{p}}%
_{0}$ can only be the global minimum of $\bar{\theta}({\mathbf{p}})$. 
\qed

\section{Conclusions} \label{sec:concl}

Building on our earlier work on suitable estimators for a wide set of Large
Deviation approximations for the probabilities of rare events, we have analyzed
the non-linear multi-dimensional optimization problems that need to be solved in
order to calculate these estimators. The special structure of the objective
function of the optimization problems at hand, suggests that a custom-made
algorithm can be much more efficient than any generic non-linear optimization
algorithm. However, proving that such an algorithm always converges is not
straightforward. In order to do so, we had to modify the simpler algorithm presented in our earlier work so that its selected step size follows a modified Armijo rule. This modified algorithm is proven to exhibit guaranteed convergence.
 
The LD estimators considered in this paper have found many applications in real-time problems such as Call Admission Control through buffer overflow probability prediction, adaptive modulation and coding with QoS constraints for transmission of data over wireless links, make-to-stock manufacturing systems and supply chains, and traffic anomaly detection in high data rate communication networks. In all the above applications, the existence of efficient and provably convergent algorithms for solving the estimation problem is a prerequisite for using the proposed LD estimators for real-time decision and control. The algorithm developed in this paper meets these requirements.

\lhead[\fancyplain{}{\bf\thepage}]{\fancyplain{}{{\sl References}}}
\bibliography{abbrev,biblio1,biblio2}

\providecommand{\bysame}{\leavevmode\hbox to3em{\hrulefill}\thinspace}
\providecommand{\MR}{\relax\ifhmode\unskip\space\fi MR }
\providecommand{\MRhref}[2]{%
  \href{http://www.ams.org/mathscinet-getitem?mr=#1}{#2}
}
\providecommand{\href}[2]{#2}
\begin{thebibliography}{BPT98b}

\bibitem[Ber95]{bert-nlp}
D.P. Bertsekas, \emph{Nonlinear programming}, Athena Scientific, Belmont, MA,
  1995.

\bibitem[BP01]{bepa1}
D.~Bertsimas and I.~Ch. Paschalidis, \emph{Probabilistic service level
  guarantees in make-to-stock manufacturing systems}, Operations Research
  \textbf{49} (2001), no.~1, 119--133.

\bibitem[BPT98a]{bpt67rev}
D.~Bertsimas, I.~Ch. Paschalidis, and J.~N. Tsitsiklis, \emph{Asymptotic buffer
  overflow probabilities in multiclass multiplexers: An optimal control
  approach}, IEEE Transactions on Automatic Control \textbf{43} (1998), no.~3,
  315--335.

\bibitem[BPT98b]{bpt5}
D.~Bertsimas, I.~Ch. Paschalidis, and J.~N. Tsitsiklis, \emph{On the large
  deviations behaviour of acyclic networks of {G/G/1} queues}, The Annals of
  Applied Probability \textbf{8} (1998), no.~4, 1027--1069.

\bibitem[Duf99]{duff1}
N.G. Duffield, \emph{A large deviation analysis of errors in measurement based
  admission control to buffered and bufferless resources}, Proceedings of the
  IEEE INFOCOM, 1999.

\bibitem[DZ98]{deze2}
A.~Dembo and O.~Zeitouni, \emph{Large deviations techniques and applications},
  2nd ed., Springer-Verlag, NY, 1998.

\bibitem[GAN91]{guahna}
R.~Gu\'{e}rin, H.~Ahmadi, and Naghshineh, \emph{Equivalent capacity and its
  applications to bandwidth allocation in high-speed networks}, IEEE Journal on
  Selected Areas in Communications \textbf{9} (1991), 968--981.

\bibitem[GH91]{gihu}
R.J. Gibbens and P.J. Hunt, \emph{Effective bandwidths for the multi-type {UAS}
  channel}, Queueing Systems \textbf{9} (1991), 17--28.

\bibitem[Gla97]{gla97}
P.~Glasserman, \emph{Bounds and asymptotics for planning critical safety
  stocks}, Operations Research \textbf{45} (1997), no.~2, 244--257.

\bibitem[GT99a]{grts1}
M.~Grossglauser and D.~Tse, \emph{A framework for robust measurement-based
  admission control}, IEEE/ACM Transactions on Networking \textbf{7} (1999),
  no.~3, 293--309.

\bibitem[GT99b]{grts2}
\bysame, \emph{A time-scale decomposition approach to measurement-based
  admission control}, Proceedings of the IEEE INFOCOM, 1999.

\bibitem[GW94]{glwh}
P.W. Glynn and W.~Whitt, \emph{Logarithmic asymptotics for steady-state tail
  probabilities in a single-server queue}, Journal of Applied Probability
  \textbf{31A} (1994), 131--156.

\bibitem[Hui88]{hui}
J.~Y. Hui, \emph{Resource allocation for broadband networks}, IEEE Journal on
  Selected Areas in Communications \textbf{6} (1988), no.~9, 1598--1608.

\bibitem[Kel91]{kel2}
F.~P. Kelly, \emph{Effective bandwidths at multi-class queues}, Queueing
  Systems \textbf{9} (1991), 5--16.

\bibitem[Kel96]{kel3}
\bysame, \emph{Notes on effective bandwidths}, Stochastic Networks: Theory and
  Applications (S.~Zachary, I.B. Ziedins, and F.P. Kelly, eds.), vol.~9, Oxford
  University Press, 1996, pp.~141--168.

\bibitem[LZG04]{liu_giann1}
Q.~Liu, S.~Zhou, and G.~B. Giannakis, \emph{Cross-layer combining of adaptive
  modulation and coding with truncated arq over wireless links}, IEEE
  Transactions on Wireless Communications \textbf{3} (2004), no.~5, 1746--1755.

\bibitem[LZG05]{liu_giann2}
\bysame, \emph{Queuing with adaptive modulation and coding over wireless links:
  cross-layer analysis and design}, IEEE Transactions on Wireless
  Communications \textbf{4} (2005), no.~3, 1142--1153.

\bibitem[LZG06]{liu_giann3}
\bysame, \emph{Cross-layer modeling of adaptive wireless link for qos support
  in heterogeneous wired-wireless networks}, Wireless Networks, Kluwer Academic
  Publishers \textbf{12} (2006), no.~4, 427--437.

\bibitem[MH88]{mur88}
D.V. Murthy and R.T. Haftka, \emph{Derivatives of eigenvalues and eigenvectors
  of a general complex matrix}, Intl. Journal for Numerical Methods in
  Engineering \textbf{26} (1988), no.~2, 293--311.

\bibitem[Pas99]{pas4rev}
I.~Ch. Paschalidis, \emph{Class-specific quality of service guarantees in
  multimedia communication networks}, Automatica, (Special Issue on Control
  Methods for Communication Networks), Anantharam and Walrand Eds., \textbf{35}
  (1999), no.~12, 1951--1968.

\bibitem[PC10]{pas-chen-tosn-10}
I.~Ch. Paschalidis and Y.~Chen, \emph{Statistical anomaly detection with sensor
  networks}, ACM Transactions on Sensor Networks \textbf{7} (2010), no.~3,
  17:1--17:23.

\bibitem[PL03]{pali1}
I.Ch. Paschalidis and Y.~Liu, \emph{Large deviations-based asymptotics for
  inventory control in supply chains}, Operations Research \textbf{51} (2003),
  no.~3, 437--460.

\bibitem[PS09]{pas-sma-ton-09}
I.~Ch. Paschalidis and G.~Smaragdakis, \emph{Spatio-temporal network anomaly
  detection by assessing deviations of empirical measures}, IEEE/ACM
  Transactions on Networking \textbf{17} (2009), no.~3, 685--697.

\bibitem[PV01]{pava01}
I.~Ch. Paschalidis and S.~Vassilaras, \emph{On the estimation of buffer
  overflow probabilities from measurements}, IEEE Transactions on Information
  Theory \textbf{47} (2001), no.~1, 178--191.

\bibitem[TZ06]{tang_zhang1}
J.~Tang and X.~Zhang, \emph{Cross-layer-model based adaptive resource
  allocation for statistical qos guarantees in mobile wireless networks},
  QShine'06 Third International Conference on Quality of Service in
  Heterogeneous Wired/Wireless Networks (Waterloo, Canada), August 2006.

\bibitem[TZ07a]{tang_zhang3}
\bysame, \emph{Cross-layer modeling for quality of service guarantees over
  wireless links}, IEEE Transactions on Wireless Communications \textbf{6}
  (2007), no.~12.

\bibitem[TZ07b]{tang_zhang2}
\bysame, \emph{Quality-of-service driven power and rate adaptation over
  wireless links}, IEEE Transactions on Wireless Communications \textbf{6}
  (2007), no.~8.

\bibitem[Var92]{var92}
H.R. Varian, \emph{Microeconomic analysis}, 3rd ed., W. W. Norton and Company,
  1992.

\bibitem[Vas10]{svas10}
S.~Vassilaras, \emph{A cross-layer optimized adaptive modulation and coding
  scheme for transmission of streaming media over wireless links}, Springer /
  ACM Wireless Networks \textbf{16} (2010), no.~4, 903--914.

\bibitem[Wil65]{wilk65}
J.H. Wilkinson, \emph{The algebraic eigenvalue problem}, Oxford University
  Press, 1965.

\bibitem[WN03]{wunegi1}
D.~Wu and R.~Negi, \emph{Effective capacity: a wireless link model for support
  of quality of service}, IEEE Transactions on Wireless Communications
  \textbf{2} (2003), no.~4, 630--643.

\bibitem[Wu03]{wu}
D.~Wu, \emph{Providing quality-of-service guarantees in wireless networks},
  Ph.D. thesis, Department of Electrical and Computer Engineering, Carnegie
  Mellon University, 2003.

\end{thebibliography}

\end{document}